# Laser-Induced Linear Electron Acceleration in Free Space

Liang Jie Wong[1,3], Kyung-Han Hong[4], Sergio Carbajo[4,5,6], Arya Fallahi[6], Marin Soljačić[2], John D. Joannopoulos[2], Franz X. Kärtner[4,5,6], and Ido Kaminer[2]

[1]Department of Mathematics, Massachusetts Institute of Technology, 77 Massachusetts Avenue, Cambridge 02139, Massachusetts, USA

[2]Department of Physics, Massachusetts Institute of Technology, 77 Massachusetts Avenue, Cambridge 02139, Massachusetts, USA

[3]Singapore Institute of Manufacturing Technology, 2 Fusionopolis Way, Innovis, Singapore 138634

[4]Department of Electrical Engineering and Computer Science and Research Laboratory of Electronics, Massachusetts Institute of Technology, 77 Massachusetts Avenue, Cambridge 02139, Massachusetts, USA

[5] Hamburg Center for Ultrafast Imaging, Luruper Chaussee 149, 22607 Hamburg, Germany

[6] Center for Free-Electron Laser Science, Deutsches Elektronen-Synchrotron, Notkestraße 85, D-22607 Hamburg, Germany

Keywords: Ultrafast lasers, ultrafast phenomena, particle acceleration

**Linear acceleration in free space is a topic that has been studied for over 20 years, and its ability to eventually produce high-quality, high energy multi-particle bunches has remained a subject of great interest. Arguments can certainly be made that such an ability is very doubtful. Nevertheless, we chose to develop an accurate and truly predictive theoretical formalism to explore this remote possibility in a computational experiment. The formalism includes exact treatment of Maxwell's equations, exact relativistic treatment of the interaction among the multiple individual particles, and exact treatment of the interaction at near and far field. Several surprising results emerged. For example, we find that 30 keV electrons (2.5% energy spread) can be accelerated to 7.7 MeV (2.5% spread) and to 205 MeV (0.25% spread) using 25 mJ and 2.5 J lasers respectively. These findings should hopefully guide and help develop compact, high-quality, ultra-relativistic electron sources, avoiding conventional limits imposed by material breakdown or structural constraints.**

The prospect of realizing high-gradient linear accelerators on small laboratory or portable scales has stimulated immense interest in laser-driven electron acceleration [1-12]. At the heart of these schemes lies the critical Lawson-Woodward theorem [13-15], which states that any laser-driven *linear* acceleration of relativistic charged particles (by a force linear in the electric field) cannot occur in free space. Thus, laser acceleration schemes typically employ assisting media like plasma [1,10,11] or nearby dielectric structures [9,16,17]. The question of whether one can achieve substantial net linear acceleration without assisting media has been studied for over 20 years using approximate treatments, and the intriguing question remains about whether this phenomenon still holds under an accurate and rigorous treatment. This question is all the more fascinating due to serious concerns that have arisen regarding the validity of the approximate treatments [18,7,19]. Here, we present exact, many-body, ab-initio simulations showing that the Lawson-Woodward theorem can be bypassed to achieve monoenergetic acceleration of a multi-electron bunch in *free space* using the longitudinal field of an ultrafast laser pulse. As examples, 30 keV electrons (2.5% energy spread) are accelerated to 7.7 MeV (2.5% spread) and 205 MeV (0.25% spread) using 25 mJ and 2.5 J lasers respectively. These findings suggest new, exciting opportunities in the development of *compact*, high-quality, ultra-relativistic electron sources that avoid conventional limits imposed by material breakdown or structural constraints.

Interest in exploring the possibility of laser-driven electron acceleration began as early as the 1970s and grew rapidly in the decades that followed, fueled by the invention of chirped pulse amplification in the 1980s [20] and a steady trend toward laser pulses of higher energies and intensities [2]. The proposal of laser-plasma acceleration [1] in 1979, for instance, was followed by a period of active research culminating in direct experimental demonstrations of the concept in the 1990s (e.g., [21]). However, it was not until 2004 that a regime for *monoenergetic* relativistic acceleration (e.g., 80-90 MeV electrons with few-percent energy spreads using 0.5 J lasers) was discovered [22-24], allowing the scheme to generate the large current beams with low energy spread necessary for many high energy electron beam applications. The broad spectrum of proposed laser-based acceleration schemes also includes inverse Čerenkov acceleration [25], inverse free-electron lasers [26,27], ionization-based acceleration [28-30,3], as well as the recently demonstrated dielectric laser accelerators [12]. All of these schemes, however, use some form of media or nearby material structures, imposing limitations in intensity and current due to practical considerations like material breakdown and damage.

Because these limits do not exist in *free space*, laser-driven acceleration in free space has the potential to take full advantage of the extremely high acceleration gradients of focused, intense laser pulses. *Linear* acceleration schemes in free space are especially noteworthy as they have advantages over non-linear (e.g., [32,33]) acceleration schemes in being less subject to transverse fields that increase the radial spread of electrons through mechanisms like ponderomotive scattering [34], and to radiation losses [35]. Due to the Lawson-Woodward theorem [13-15], however, one would expect *functional* linear acceleration – involving the mono-energetic acceleration of multiple charged particles occupying a finite volume in space – to be impossible unless physical media or nearby material structures are present. Despite early indications [36-40] that linear acceleration in free space is possible for a single, on-axis particle – see Supplementary Information (SI) Section S1 for a discussion – it is not clear whether an actual electron pulse composed of multiple electrons can be accelerated in a stable and controlled way. Instead, one might expect that the electron pulse would acquire a large energy variance (e.g., due to inter-electron repulsion), resulting in a distribution of accelerated and decelerated electrons such that the average net acceleration is negligible. This has never been rigorously tested since there exists no study on laser-driven linear acceleration that takes into account the inter-particle interactions of a multi-electron bunch. Importantly, it was also suspected that any predicted linear acceleration (e.g., [36]) was an artifact of certain approximations, such as the paraxial approximation [18,7]. For the above reasons, the possibility of linear acceleration in unbounded free space has remained an open subject of great interest.

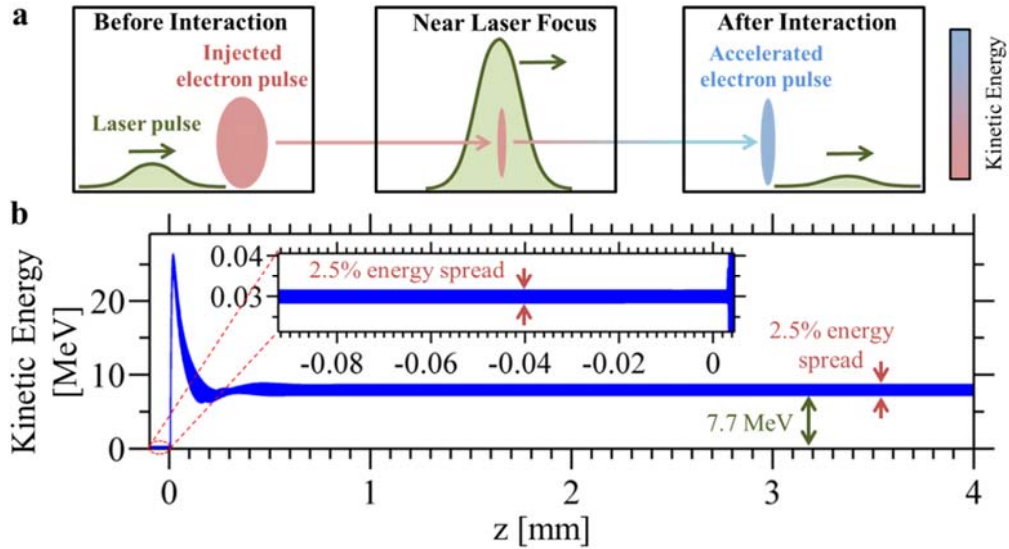

**Figure 1. Linear electron acceleration in unbounded free space by an ultrafast radially-polarized laser pulse, a process illustrated schematically in (a). As an example, (b) shows the net acceleration to a final energy of 7.7 MeV of a 30 keV electron pulse with charge -0.2 fC, by a 25 mJ, 3 fs laser pulse of wavelength 0.8 μm focused to a waist radius of 1.6 μm. The initial electrons are randomly distributed in a sphere of diameter 1 μm. More details of the interaction in (b) are given in Fig.2.**

To answer these fundamental questions, we develop an exact, multi-particle electrodynamics simulation tool in which laser-driven linear acceleration is treated *without any approximations* by treating multiple individual particles (see SI Section S2). This provides us with new predictive powers that allow us to explore *monoenergetic* net acceleration of a multi-electron bunch with a radially-polarized laser pulse, and demonstrate using a rigorous theoretical formalism that functional linear acceleration is possible in free space. Our findings suggest that high-gradient linear acceleration of electron pulses containing a large number of particles can be achieved with an energy spread comparable to or even smaller than current state-of-the-art acceleration techniques. Our scheme only requires engineering the spatiotemporal structure of light in unbounded free space, avoiding the use of media (e.g., gas or plasma), nearby material boundaries, or static fields. Our findings thus constitute the design of the first free-space linear acceleration scheme for multi-electron bunches.

Fig. 1 shows the acceleration of 30 keV electrons (2.5% spread) to 7.7 MeV (2.5% energy spread) with a 25 mJ pulse. The scheme we study uses the ultrafast radially-polarized laser pulse [41,42], an attractive candidate for electron acceleration due to the ability of its transverse fields to confine electrons to the axis exactly where the longitudinal electric field peaks and linear acceleration is most effective. We use a carrier wavelength of 0.8 μm, full-width-at-half-maximum (FWHM) pulse duration 3 fs (6 fs is studied in SI Section 6), and waist radii (second irradiance moment at focal plane) ranging from $w_0$ = 0.8 μm to 5.0 μm. It has been shown that such tightly-focused radially-polarized laser beams can be created in practice using parabolic mirrors of high numerical aperture [43], assisted by wavefront correction with a deformable mirror [44]. Electrons can be injected into the focused region through a small hole on the parabolic mirror. Very recently, millijoule-level, few-cycle pulsed radially-polarized laser beams capable of reaching intensities above $10^{19}$ W/cm$^2$ with kilohertz repetition rates have been experimentally demonstrated [42].

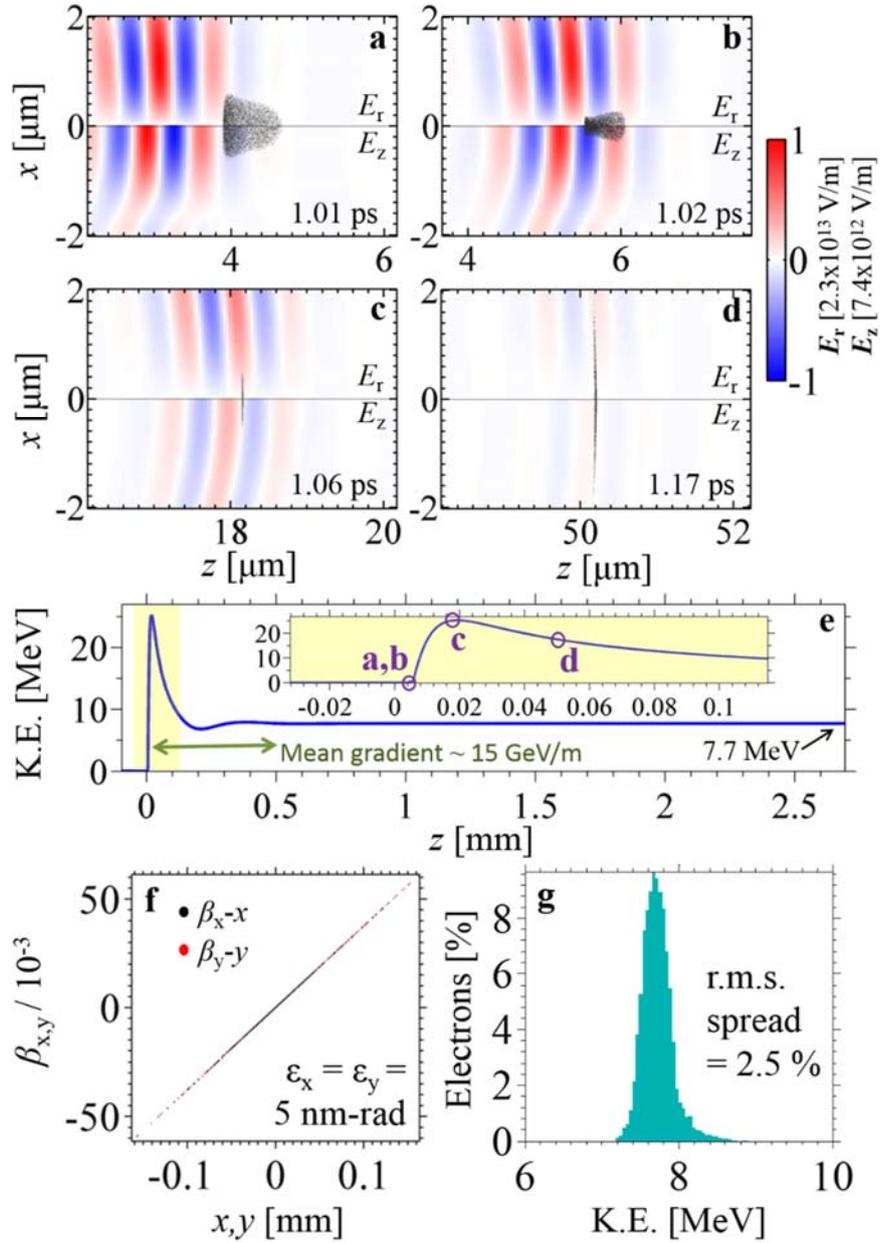

**Figure 2. Monoenergetic, relativistic electrons from laser-driven linear acceleration in free space.** (a)-(d) depict the behavior of the optical and electron pulses at various times during the laser-electron interaction. These instants are marked with circles in (e), which show the evolution of the electron pulse's mean kinetic energy as a function of distance (laser focus at $z=0$). The final (f) normalized trace-space emittance and (g) energy distribution describe a relativistic, high-quality and quasi-monoenergetic electron pulse. The laser and electron pulse parameters from Fig. 1 were used here. Although the electron pulse eventually acquires a relatively large transverse size (see (f)), its low trace-space emittance implies that it is readily re-compressed with appropriate focusing elements (e.g.: magnetic solenoid).

Figs. 2(a)-(d) capture the laser-driven linear acceleration process at various instants (see the Supplementary Video for an animation). At initial time $t = 0$, the injected electrons travel in the complete absence of electromagnetic fields. At $t \approx 1$ ps, the optical pulse overtakes the focused laser pulse close to the laser beam focus $z = 0$ (Fig. 2(a)), where the superluminal optical phase velocity causes electrons to slip rapidly through accelerating and decelerating cycles (Fig. 2(b)). As the laser beam diverges, the superluminal phase velocity decreases towards the speed of light c even as the electrons, caught in an accelerating cycle, accelerate towards c, leading to a period of sustained acceleration. At some point, the electrons slip into a decelerating cycle and start losing energy (Fig. 2(c)). Due to the optical beam divergence and finite pulse duration, however, the electrons can still retain a substantial amount of the energy gained after escaping the influence of the optical pulse. The final electron pulse of -0.2 fC – which travels once more in field-free vacuum (Fig.1a right panel) – is quasi-monoenergetic with a mean energy of 7.7 MeV, an energy spread of 2.5% and normalized trace-space emittances [45] of 5 nm-rad (Fig. 2(d)-(f)). Note that at 7.7 MeV, this corresponds to an unnormalized trace-space emittance of about 0.31 nm-rad, smaller than the initial trace-space emittance of 1 nm-rad. The initial electron bunch is focused to a diameter of 1 μm in each dimension (10 fs pulse duration).

The generation of initial sub-relativistic electron pulses of few-fC charge, sub-wavelength transverse dimensions and few-femtosecond durations is known to be achievable, e.g. by ionizing a low-density gas with a laser beam [28,46], all-optical compression techniques [47,48] or photoemission from sharp metal tips [49,50]. The modest amount of charge considered in our scheme may be scaled up by employing a high-repetition-rate electron source and an external cavity to recycle the laser pulse, leading to average currents as high as 2 μA for a 1 GHz repetition rate. Relaxing the requirements on the output electron pulse, employing longer laser wavelengths, and increasing the laser pulse energy (see Fig. 3) are also ways of increasing the amount of charge that can be accelerated. We chose a uniform spheroidal distribution for our initial electron pulse due to the well-behaved nature of such distributions [51], but simulations with other distributions (e.g., a truncated Gaussian distribution) give practically identical results.

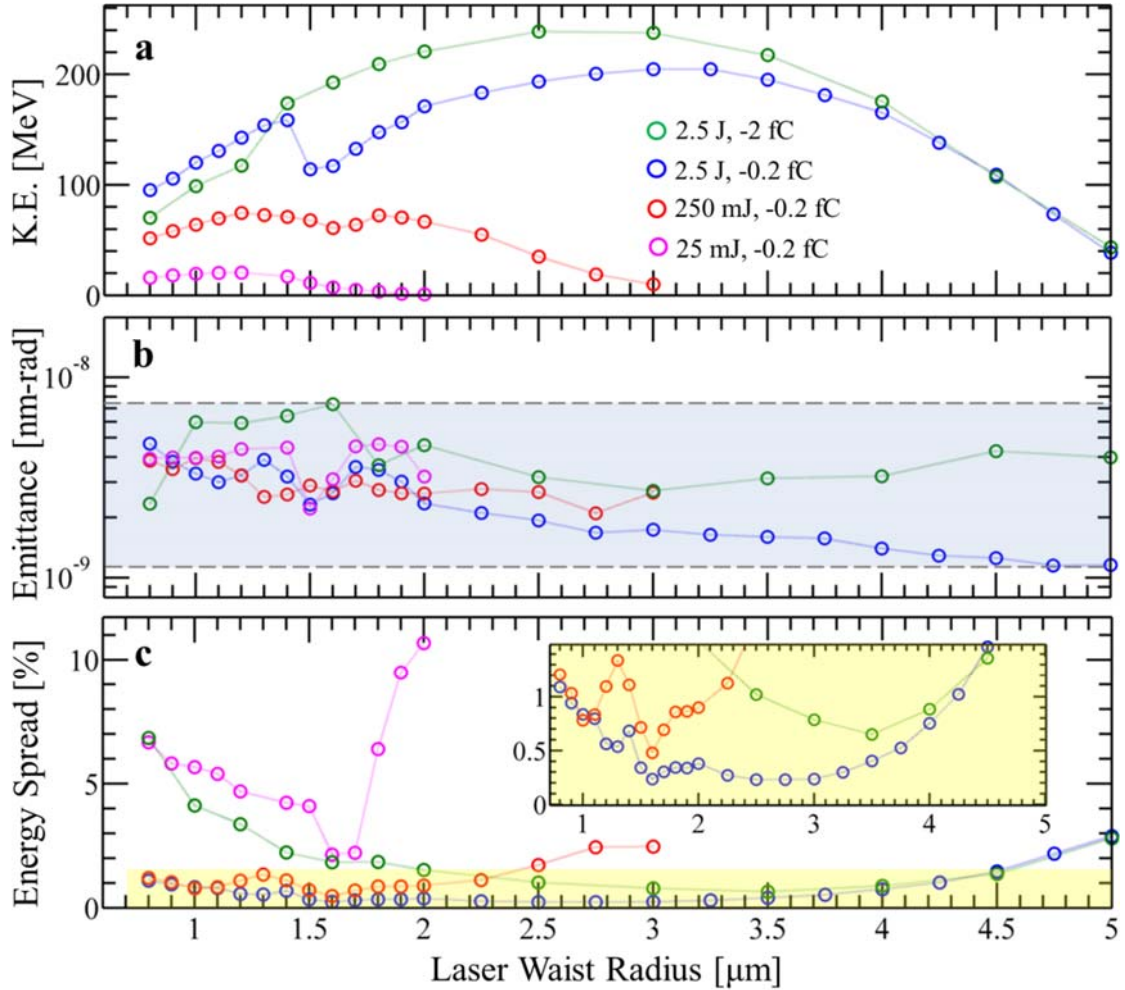

**Figure 3. Characteristics of the accelerated electron pulse, showing the insensitivity of laser-driven linear acceleration in free space to a wide range of parameter choices. The final (a) mean kinetic energy, (b) normalized trace-space emittance, and (c) energy spread are shown as a function of laser waist radius for various values of laser pulse energy and electron pulse charge. The shaded region between the dashed lines in (b) is to highlight the fact that for a wide range of parameters, the final emittance falls in the nm-rad range. We obtain each point by optimizing over the optical carrier phase and the relative displacement between electron and laser focal positions based on our FOM. (Dotted lines are included only as a visual guide).**

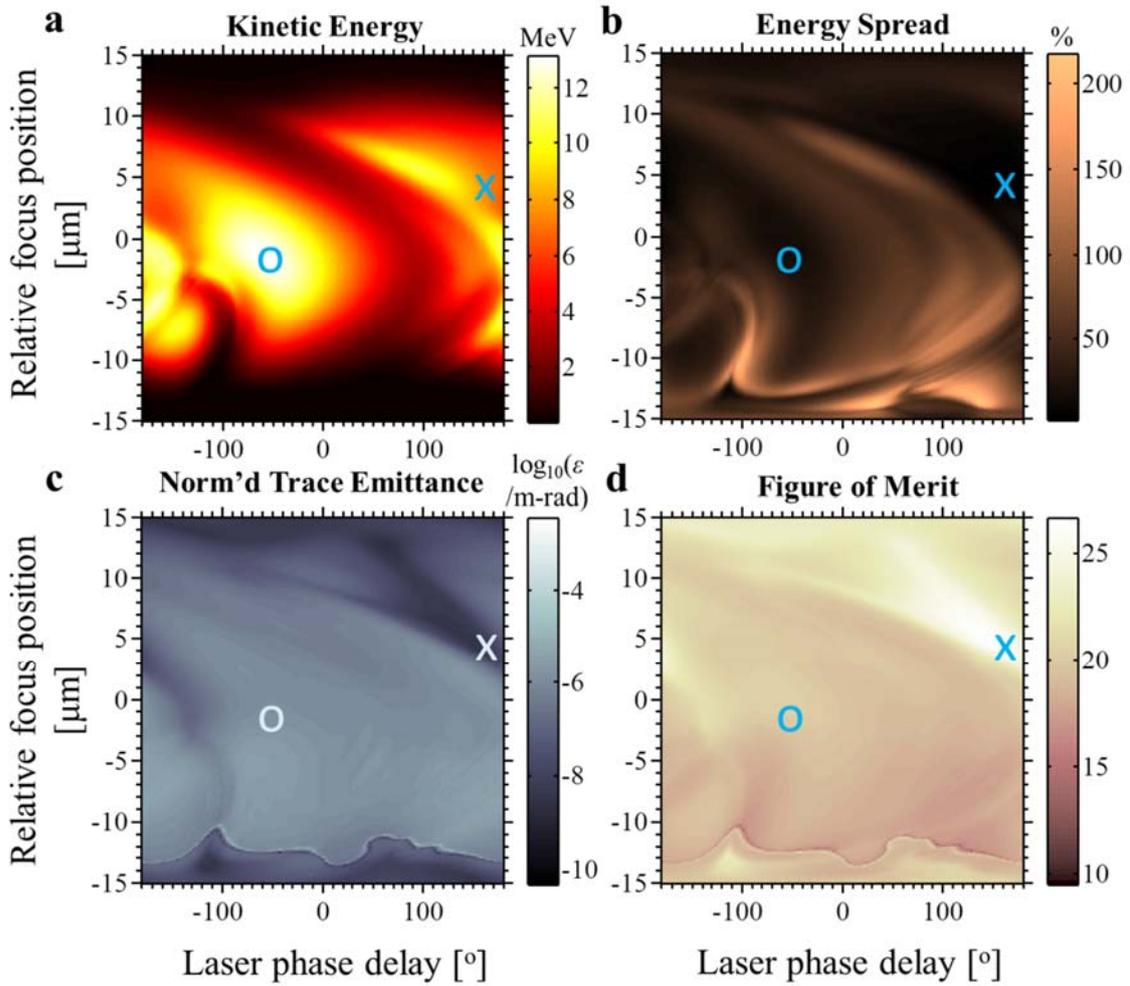

**Figure 4. Determining optimal parameters for free space linear acceleration.** The panels show large regimes of high-quality acceleration, allowing optimization of the scheme under different figures-of-merit (FOMs). (a) Mean kinetic energy, (b) energy spread, (c) normalized trace-space emittance and (d) FOM of the final electron bunch after acceleration as a function of the optical carrier phase and the position of the electron beam focus. Along the vertical axes, we vary the position of the interaction point (where the electron pulse reaches its focus) relative to the spatial focus of the laser. Along the horizontal axes, we vary the laser carrier phase across all possible phase delays. The laser pulse and initial electron pulse are identical to those used in Figs. 1 and 2. In every case, the electron pulse is designed to reach its temporal focus and its spatial focus simultaneously. The temporal focus of the laser pulse is synchronized to coincide with that of the electron pulse. All properties are recorded long after the laser-electron interaction has ceased. The location corresponding to maximum net acceleration is marked with a circle ('o'), whereas the location of the optimal solution according to our FOM (which takes emittance and energy spread into account; see text) is marked with a cross ('x'). This optimal solution corresponds to the results in Figs. 1 and 2.

Figure 3 shows the results of the scheme for larger laser pulse energies. For instance, we will see that final energies of 61 MeV (0.5% energy spread) and 205 MeV (0.25% energy spread) can be attained with 250 mJ and 2.5 J laser pulses respectively (this is further discussed in SI Sections S3 and S4). The scaling of the maximum kinetic energy gain $\Delta U$ roughly obeys the $\Delta U \propto P^{1/2}$ law ($P$ being peak pulse power), corresponding to a linear scaling in peak electric field, which confirms that the acceleration mechanism is linear.

Both Figs. 3 and 4 emphasize the stability of our scheme by showing that small changes in parameters do not affect the results considerably. For example, a displacement from the optimal point of 1 μm in focal position and 10° in phase delay in Fig. 4 would only degrade the peak acceleration by 1% and the emittance by 5%. Our acceleration scheme is also robust to variations in initial electron energy spread (SI Section S5) and pulse duration (SI Section S6), giving monoenergetic, high-emittance acceleration of multi-electron pulses even with initial electron energy spreads as large as 40% or with a laser pulse duration of 6 fs.

The optimal combination of parameters for the linear acceleration scheme is ultimately subjective since different applications have different requirements for energy, energy spread and emittance. As Fig. 4 illustrates, the maximum energy gain (marked 'o') does not in general correspond to the best normalized trace-space emittances and energy spread. Heuristically, we have found that a useful figure-of-merit (FOM) in determining arguably optimal conditions is $\mathrm{FOM} \equiv (\langle\gamma\rangle-1)^5 / (\varepsilon_x \varepsilon_y \Delta\gamma^4)$, where $\gamma$ is the relativistic Lorentz factor, $\Delta\gamma$ the root-mean-square spread in $\gamma$, and $\varepsilon_{x,y}$ the normalized trace-space emittances of the final electron pulse. The parameters used in Figs. 1 and 2 were obtained using this FOM (which maximum is marked 'x' in Fig. 4). Note that the point of maximum energy gain is not located at the laser focus and zero carrier-envelope phase, since the focus has both the strongest electric field amplitude, which favors large acceleration, as well as the most superluminal phase velocity, which encourages phase slippage and thus hinders acceleration. It is likely that optimizing over even more degrees of freedom can further improve the properties of the accelerated pulse. These degrees of freedom include the optical pulse duration and the spatiotemporal structure of the laser pulse, which one can control with external optical components.

Let us now address an intriguing question: How is it that linear acceleration in free space has never been observed in laser-driven acceleration experiments, although this is possible over a relatively wide range of parameters? The immediate reason, which also highlights an important requirement in our scheme, is the use of ultrashort pulses that are also of significant intensities (importantly, this requirement falls within the reach of current experimental capabilities). Our findings thus strongly motivate the development of ultra-intense, few-cycle and even sub-cycle laser pulses, as well as better control over the polarization and phase of such pulses. We note that single-cycle pulses [52] and radially-polarized few-cycle pulses [42] have been experimentally demonstrated in standard laboratories.

The results described in this work give insights beyond our range of parameters and our choice of laser wavelength, because solutions to the Newton-Lorentz equation of motion and Maxwell's equations are scale invariant in the absence of space charge. Except for a scaling factor in laser and electron pulse parameters, we expect our results to remain relevant at other wavelengths, modulo some correction terms due to space charge. Terahertz sources have observed a steady trend toward pulses of higher intensity and energy [53-55], and could be attractive alternatives to optical or infrared sources due to the larger amount of charge that can be accommodated at terahertz wavelengths.

Our results also strongly suggest the viability of linearly accelerating other types of charged particles in free space (e.g., protons and ions, for applications like hadron therapy in cancer treatment and lithography by ion beam milling). Generally, the charged particles may be externally injected, and do not have to be introduced by methods like ionization [56] that require media or material structures near the laser focus. Weaker inter-particle interactions between heavier charged particles may enable higher current and even more impressive performance of the acceleration scheme. Additionally, note that the linear acceleration gradient is given by electric field amplitude $E$, whereas the ponderomotive acceleration gradient is given by $qE^2/(2\omega\gamma mc)$ [57], where $q$ and $m$ are respectively the particle's charge and rest mass, $\omega$ is the central angular frequency of the laser, $\gamma$ is the relativistic Lorentz factor, and c is the speed of light in free space. As a result, heavier charged particles (with larger $m$) are likely to be more strongly favored by linear acceleration.

In conclusion, we have shown that net energy transfer between laser and multi-electron pulses via linear forces can be achieved by engineering the spatiotemporal structure of light in unbounded free space. Our findings motivate the development of ultra-intense, few-cycle and even sub-cycle laser pulses, as well as better control over the polarization and phase of such pulses. Rapid technological advances in engineering arbitrary wavefronts and polarizations [58], together with emerging techniques for precise structuring of electron pulses [59,60], create a wealth of opportunities that will push the limits of particle acceleration to ever-higher energies.

# Laser-Induced Linear Electron Acceleration in Free Space

Liang Jie Wong, Kyung-Han Hong, Sergio Carbajo, Arya Fallahi, Marin Soljačić,

John D. Joannopoulos, Franz X. Kärtner, and Ido Kaminer

## Supplementary Information

Contents

## S1  Linear acceleration in free space under the Lawson-Woodward theorem

The notion of a particle gaining net energy from the longitudinal component of a purely-propagating electromagnetic wave in unbounded free space is a counter-intuitive one. It would appear that since such a particle sees as many accelerating cycles as decelerating cycles, the net energy gain should be approximately zero. This is indeed true if the particle maintains an approximately constant velocity (say, a velocity close to c, the speed of light in vacuum) throughout the laser-particle interaction. However, when the velocity varies greatly – as is the case when the laser is powerful enough to modulate the particle's velocity between relativistic and non-relativistic regimes [S1, S2] – the energy change due to adjacent cycles can be very different and substantial net acceleration becomes possible. At first glance, this phenomenon seems to contradict the Lawson-Woodward theorem, as derived in [S3], which states that a particle that remains highly relativistic throughout its interaction with the longitudinal electric field of a propagating wave gains an insubstantial amount of energy. (We note in passing that the proof in [S3] treats continuous-wave electromagnetic beams but is readily generalized to the case of a pulsed electromagnetic field).

The proof in [S3] leads to the exact mathematical condition:

$$\forall t, \quad \left|\frac{dz}{dt}\right| = c \quad \Rightarrow \quad \Delta U = 0, \tag{S1}$$

where $\Delta U$ is the net energy gain in unbounded free space from the purely-linear acceleration process, and $z$ is the particle's position at time $t$. Although $|dz/dt| = c$ is a physical impossibility for a massive particle, it is a reasonable approximation for the speed of a highly-relativistic particle, which travels at $|dz/dt| \approx c$. Reversing (S1) enables us to obtain the condition for the substantial net linear acceleration that we seek:

$$\Delta U \neq 0 \quad \Rightarrow \quad \exists t, \left|\frac{dz}{dt}\right| \neq c. \tag{S2}$$

In other words, it is possible to obtain net linear acceleration in unbounded free space if there exists a time during the interaction where the particle is non-relativistic [S1, S2]. More generally, for substantial net linear acceleration to take place, the laser pulse must be powerful enough to take the particle between relativistic and non-relativistic regimes, in the original particle's rest frame. This observation has been used to obtain an analytical approximation for the onset of substantial acceleration, marked by the boundary between the regimes of substantial and insubstantial net particle acceleration in free space (assuming the purely-longitudinal on-axis electric field of a propagating wave is involved [S1]).

This threshold, defined in terms of normalized vector potential $a$, generalizes the $|a| \approx 1$ threshold [S4] often invoked to distinguish between relativistic and non-relativistic laser-particle interaction regimes. The conventional threshold assumes that the electron velocity in the direction of the electric field is zero on average, which is a reasonable assumption in most cases since the electric field is perpendicular to the direction of net electron motion. However, the scenarios studied here involve longitudinal fields that are parallel to the direction of electron motion. This requires a more generalized threshold expression, which is given by:

$$\left[\left|\frac{a}{\gamma_0(1-v_0/v_{ph})}\right|\right]_{max} \approx 1, \quad (S3)$$

where $v_0$ is the initial, z-directed velocity of the injected particle (initially in field-free vacuum), and $\gamma_0 = (1-v_0^2/c^2)^{-1/2}$ is the corresponding initial Lorentz factor. The normalized vector potential $a \equiv qE_z/mc\omega$, where $E_z$ is the amplitude of the longitudinal electric field. $v_{ph}$ is the phase velocity of the electromagnetic wave, $\omega$ is the central frequency of the laser pulse, and $q$ and $m$ are the particle's charge and mass respectively. Naturally, both the normalized vector potential $a$ and the phase velocity $v_{ph}$ vary along a focused laser beam. The $[\cdot]_{max}$ in (S3) instructs us to use the local value of $a$ and $v_{ph}$ at the point along the beam where the expression in the square parentheses is maximum. In many cases, this occurs at the laser beam focus.

S1.1 The acceleration threshold for a radially-polarized laser pulse

In this section, we give an example showing the effectiveness of (S3) in predicting the boundary between regimes of substantial and insubstantial linear acceleration in free space. We simulate the exact electrodynamics of the electron interacting with the radially-polarized laser pulse for a range of initial electron energies and pulse energies. As in all cases throughout this work, the simulations begin and end with the electron in (effectively) field-free vacuum. A color map plot of the final electron momenta shows a clear borderline between the regimes of significant and insignificant acceleration, which we compare to the prediction of Eq. (S3). Our procedure is similar to that in [S1], except that here we model the radially-polarized laser pulse using an exact solution to Maxwell's equations. Fig. S1 shows a color map of the final momentum of a single, on-axis electron interacting with the radially-polarized laser pulse. The color map – which is obtained from exact numerical calculations – presents the optimized final momenta as a function of the initial electron momentum and the pulse energy. (For each combination of initial electron momentum and pulse energy values, the optimization is performed over relative focal position — namely, the position of the electron when the laser pulse arrives at the beam focus — and the carrier envelope phase.) To facilitate comparison between cases with different initial electron momenta, the final momenta in Fig. S1 are given in the initial rest frame of the electron. We have assumed a laser pulse of wavelength 0.8 μm, pulse duration 6 fs (intensity FWHM) and waist radius 1.2 μm (second irradiance moment).

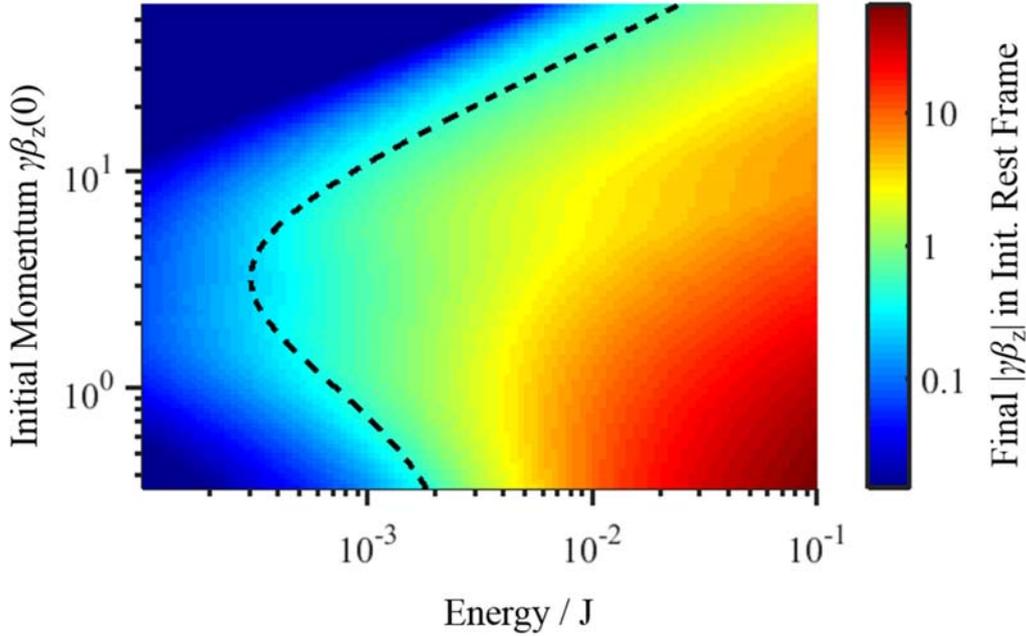

**Figure S1. Color map showing maximum final momentum of an on-axis electron interacting with a radially-polarized laser pulse. The final momentum is presented as a function of the initial electron momentum (y-axis) and the pulse energy (x-axis). To compare the net acceleration across cases of different initial electron momentum, this final momentum is presented in the initial rest frame of the electron. Momentum is presented in dimensionless units (normalized by mc). The dotted black curve shows the threshold of substantial acceleration as predicted by the analytical formula (S3), demonstrating the accuracy of this formula.**

Fig. S1 also plots our analytical formula Eq. (S3) predicting the onset of substantial net energy gain. The curve matches very well with the boundary between the regime of insubstantial and substantial net linear acceleration of charged particles in unbounded free space. Fig. S1 thus demonstrates the effectiveness of the analytical approach leading to Eq. (S3) [S1]. This procedure also verifies the hypothesis propounded in [S1] that for substantial net linear acceleration to take place, the laser pulse must be powerful enough to take the particle between relativistic and non-relativistic regimes, in the original particle's rest frame.

## S2     Theory and numerical implementation of electrodynamics and radiation reaction

The numerical results in this work were obtained through exact, ab-initio simulations of laser-electron interactions in free space. Since our simulations solve the exact electrodynamics problem, they include not only the standard Newton-Lorentz dynamics and the Coulomb repulsive force (a.k.a., space charge), but also the far-fields of charged particle motion and radiation reaction. Using these simulations, we demonstrate not only high quality linear acceleration of multi-electron bunches, but also that the resulting acceleration is not attributable to any type of inter-particle interaction or radiation reaction, and is stable for a range of parameters.

More specifically, our model includes 1) an exact, non-paraxial solution of Maxwell's equations for a pulsed, focused radially-polarized laser beam containing no static frequency components, 2) inter-particle interactions modeled using the Liénard-Wiechert potentials, which correspond to exact solutions for both the near and far fields of moving charged particles, and 3) radiation reaction, the recoil a particle experiences due to radiation losses (and which effect we ascertain to be negligible in our parameter space of interest, as we note in Section S4). Our numerical program uses a particle-tracking algorithm based on an adaptive-step fifth-order Runge-Kutta algorithm [S5]. For reasons associated with performance, it is customary in some cases to assign more or less than one electron to each simulation particle. However, this approach can potentially create undesirable artifacts due to the presence of the far-field, which can cause coherent addition effects to become important. Therefore, we represent each electron using exactly one simulation particle (e.g., employing 12,500 simulation particles to simulate -2 fC of electrons), averaging over multiple runs to obtain the results presented in statistical plots like Fig. 3. The averaging is necessary to ensure that any feature in the bunch evolution is not due to a coincidence of the initial random distribution. Notwithstanding the above considerations, it is worth noting that since the influence of the far-field is typically insignificant in our regime of interest here, assigning more or less than one electron to each simulation particle (keeping total charge constant) does not lead to any significant change in the results, as long as the number of simulation particles is large enough (usually > 1000).

Although we employ an exact model, we have verified that in our regime of interest, the radiation reaction effect is negligible. In addition, the exact computation of retarded time in the Liénard-Wiechert potentials can be replaced by an approximate retarded time – obtained by assuming each particle has always been traveling at the velocity it possesses in its current time step – with negligible implications for the near field in inter-particle interactions (see Section S4).

The effect of the far field from charged particle motion begins to have a significant effect on the quality of the electron pulse only after a time that is significantly longer than the duration of the laser-electron interaction. These observations imply that in many cases, our results can be reproduced to a fair degree of accuracy under the assumptions of standard particle-tracking solvers (e.g. the General Particle Tracer [S6]). Note that standard particle-tracking solvers typically do not take the far field and radiation reaction into account. When computing the effect of inter-particle interactions, standard particle-tracking solvers also typically make the assumption that each particle has always been traveling at its instantaneous velocity at each time step. Our simulations make none of these assumptions, and instead solves the exact physical problem.

In the following sub-sections, we elaborate on the underlying theory, numerical implementation, and practical implications of our model.

S2.1  Electrodynamics in free space

The electrodynamics of a charged particle cloud interacting with an electromagnetic field are governed by the Newton-Lorentz equation of motion [S7]. When the effect of radiation reaction [S15] is included, the dynamics of each particle is given by

$$K^\mu = q\eta_\nu F^{\mu\nu} + R^\mu, \tag{S4}$$

where $K^\mu = dp^\mu/d\tau$, $p^\mu = m\eta^\mu$ is the 4-momentum, $\tau$ the proper time, $\eta = \{\gamma c, \gamma v_x, \gamma v_y, \gamma v_z\}$ the 4-velocity, $\gamma = (1-\beta^2)^{-1/2}$ the Lorentz factor (with $\beta = |\boldsymbol{\beta}|$ and $\boldsymbol{\beta} = \mathbf{v}/c$), $q$ the particle's charge, $m$ the particle's mass, $\mu_0$ the permeability of free space, and c the speed of light in free space. The metric used is $\{-,+,+,+\}$. $R^\mu$ captures the effect of radiation reaction, namely, the recoil the charged particle experiences as it loses energy by radiation. The electromagnetic tensor $F^{\mu\nu}$ is given in matrix form as (with rows indexed by $\mu$ and columns by $\nu$)

$$F = \begin{bmatrix} 0 & E_x/c & E_y/c & E_z/c \\ -E_x/c & 0 & B_z & -B_y \\ -E_y/c & -B_z & 0 & B_x \\ -E_z/c & B_y & -B_x & 0 \end{bmatrix}, \tag{S5}$$

where the electric and magnetic components comprise the driving electromagnetic field as well as fields due to the presence of other charged particles in the cloud.

S2.2  Non-paraxial, analytical model of a radially-polarized laser pulse

An exact, non-paraxial, analytical solution of Maxwell's equations for a pulsed, focused radially-polarized Laguerre-Gaussian TM10 mode in free space [S8] is given by

$$\mathbf{E} = \left[\hat{\mathbf{r}}\frac{\partial^2}{\partial r \partial z} - \hat{\mathbf{z}}\frac{1}{r}\frac{\partial}{\partial r}\left(r\frac{\partial}{\partial r}\right)\right]\Psi$$

$$\mathbf{B} = -\hat{\phi}\frac{1}{c^2}\frac{\partial^2}{\partial r \partial t}\Psi$$

(S6)

using the Hertz vector potential

$$\Psi(t) = \mathrm{Re}\left[\frac{C_0}{R'}\left(f_+^{-s-1} - f_-^{-s-1}\right)\right],$$

(S7)

where $f_\pm = 1 - i/s(\omega_0 t \pm k_0 R' + ik_0 a)$, $R' = \sqrt{x^2 + y^2 + (z+ia)^2}$, and $C_0$ is some complex constant. The peak angular frequency of the pulse spectrum is $\omega_0 = k_0 c = 2\pi c/\lambda_0$. The degree of focusing (or beam waist radius $w_0$) and the pulse duration $\tau$ are set through parameters $a$ and $s$. Good analytical approximations [S9, S8] relating $a$ and $s$ to $w_0$ and $\tau$ are used as starting points for numerical iterations that further refine $a$ and $s$ for user-specified $w_0$ and $\tau$. The Fourier transform of the square-bracketed expression in (S7) is proportional to

$$\widetilde{\Psi}(\omega) = \frac{\sin(kR')}{R'\exp(ka)}\left[\omega^s \exp\left(-\frac{s\omega}{\omega_0}\right)\theta(\omega)\right],$$

(S8)

where $k = \omega/c = 2\pi/\lambda$, $\lambda$ is the vacuum wavelength, and $\theta(\cdot)$ is the Heaviside step function. In (S8), the pre-factor on the right-hand-side is associated with the diverging beam, whereas the square-bracketed expression – the Poisson spectrum – is associated with the pulse envelope. Eq. (S8) is an exact solution of the Helmholtz equation at any $\omega$. Note that $\widetilde{\Psi}(\omega = 0) = 0$, implying that (S6)-(S7) contain no static frequency components and thus describe an electromagnetic field that is <u>purely propagating</u>. We emphasize that equations (S6)-(S8) are exact solutions for broadband as well as narrow-band electromagnetic pulses.

Throughout this work, beam waist radius $w_0$, laser pulse energy $U$ and pulse duration $\tau$ obey the following definitions (Eqs. S6, S7, and S8 respectively), which can be computed via numerical cubature [S9]:

$$w_0 = \sqrt{\frac{\iint r^2 S_z' \, dxdy}{\iint S_z' \, dxdy}},$$

(S9)

which is equivalently the second irradiance moment of the pulse at the focal plane and pulse peak. The double integrals are over the entire focal plane and $S_z'$ is the z-directed Poynting vector component $S_z \equiv \vec{E} \times \vec{H} \cdot \hat{z}$ evaluated at the focal plane, pulse peak and carrier amplitude of the

TM10 mode. This definition of beam waist is motivated by the fact that in the paraxial, continuous-wave (CW) limit, we obtain fields $\sim \exp(-r^2/w_0^2)$. The pulse energy is defined as

$$U = \iiint S_z \, \mathrm{d}x\mathrm{d}y\mathrm{d}t, \qquad (S10)$$

and the pulse duration as

$$\tau = 2\sqrt{2\ln 2}\sqrt{\frac{\iiint t^2 S_z \, \mathrm{d}x\mathrm{d}y\mathrm{d}t}{U}}, \qquad (S11)$$

where the triple integrals are over the entire focal plane and temporal axis. This definition of pulse duration is motivated by the fact that in the many-cycle limit, the Poisson spectrum in (S8) approaches a Gaussian spectrum and $\tau$ then corresponds to the full-width-half-maximum (FWHM) duration of the Gaussian pulse.

It is important to note that the exact solution to the radially-polarized laser pulse is one that possesses a finite volume in four-dimensional space time. Figure S2 shows the transverse profile decay of the exact solution given by Eqs. (S6) and (S7) at the focal plane. The decay approaches a Gaussian rate of decay in the limit of large pulse durations, but nevertheless remains very rapid even for a pulse duration as short as 3 fs.

In this work, we use (S6)-(S7) to model laser pulses of carrier wavelength 0.8 μm and pulse duration 3 fs focused down to as small a waist as $w_0$ = 0.8 μm. It has been shown that such tightly-focused radially-polarized laser beams can be created in practice using parabolic mirrors of high numerical aperture [S10], assisted by wavefront correction with a deformable mirror [S11]. Electrons can be injected into the focused region through a small hole on the parabolic mirror. Very recently, millijoule-level, few-cycle radially-polarized laser pulses capable of reaching intensities above $10^{19}$ W/cm$^2$ with kilohertz repetition rates have been experimentally demonstrated [S12].

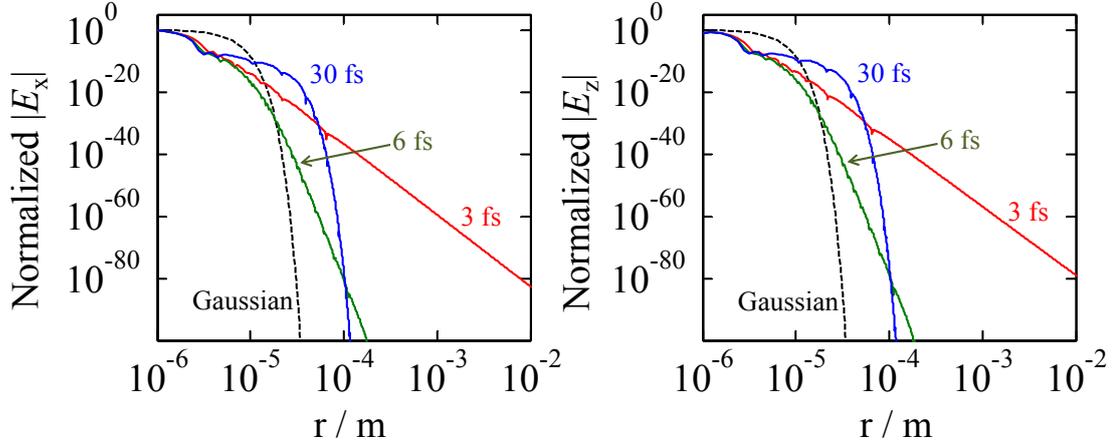

**Figure S2. Transverse decay of an exact solution for the radially-polarized laser pulse.** The normalized electric fields (a) $E_x$ and (b) $E_z$ corresponding to the exact solution Eqs. (S6) and (S7) are shown for different pulse durations (curve labels). The decay of a Gaussian profile of standard deviation equal to the beam waist radius (1.6 μm) is shown for reference. The transverse decay of the exact solution thus approaches Gaussian decay in the limit of large pulse durations, and remains very rapid even for a pulse duration as short as 3 fs.

S2.3  Inter-particle interactions and the implementation of retarded time calculations

Charged particles interact with one another through both their near fields and far fields (radiation fields). Due to the time it takes for electromagnetic effects to propagate, the force that charged particle $i$ experiences due to charged particle $j$ at time $t$ is related to the properties of particle $j$ not at time $t$, but at the retarded time $t'$, given by

$$t' = t - \frac{R}{c}, \tag{S12}$$

where $R = |\mathbf{R}|$, with $\mathbf{R} = \mathbf{r}_i(t) - \mathbf{r}_j(t')$ and $\mathbf{r}_i(t)$ the position of particle $i$ at time $t$. The total electromagnetic fields that particle $i$ experiences due to particle $j$ at time $t$ are obtained from the Liénard-Wiechert potentials as [S7]

$$\mathbf{E} = \frac{q}{4\pi\varepsilon_0 (1 - \hat{\mathbf{n}} \cdot \boldsymbol{\beta})^3 R} \left\{ \frac{\hat{\mathbf{n}} - \boldsymbol{\beta}}{R\gamma^2} + \frac{\hat{\mathbf{n}} \times [(\hat{\mathbf{n}} - \boldsymbol{\beta}) \times \dot{\boldsymbol{\beta}}]}{c} \right\},$$

$$\mathbf{B} = \frac{\hat{\mathbf{n}} \times \mathbf{E}}{c}, \tag{S13}$$

where $\dot{\boldsymbol{\beta}} = \partial\boldsymbol{\beta}/\partial t'$ and $\varepsilon_0$ is the permittivity of free space. All right-hand-side variables belong to particle $j$ and are evaluated at retarded time $t'$, except for **E**, $\hat{\mathbf{n}} = \mathbf{R}/R$ and **R** (which involve both particles and also current time $t$). The first term on the right-hand-side of the first equation in (S13) corresponds to the near field, whereas the second term corresponds to the radiated far field.

To solve (S4) numerically in the presence of inter-particle interactions (S13), we store a history of **r**, $\gamma\boldsymbol{\beta}$ and d($\gamma\boldsymbol{\beta}$)/dt for every particle. At each time step, we use cubic spline interpolation [S5] to determine the retarded times (S12) and the corresponding parameters to be used in (S13) for each particle with respect to every other particle. In our regime of interest, we find that a 2048-element-long history with storage times spaced about $0.05\lambda_0/c$ apart is usually sufficient to produce results within 1% of the exact values. This algorithm has a time complexity of O($N^2\log P$), where $N$ is the number of particles and $P$ the length of the stored history (the log($P$) factor arises from the fact that we use a binary search algorithm to locate the retarded times).

For comparison, we also perform simulations using the approximations of standard particle-tracking solvers. Since our simulations make no such approximations this is not necessary, but it is insightful for computational reasons since it significantly speeds up the computation. To speed up the computation of inter-particle forces, standard particle-tracking solvers (e.g. the General Particle Tracer [S6]) often ignore the far field component in (S13). Furthermore, they assume at every step of the ordinary differential equation solver that each particle has always been traveling at its current velocity. This is mathematically equivalent to the procedure of Lorentz-boosting the static Coulomb field of each charged particle from its instantaneous rest frame to the lab frame. As a result, (S13) simplifies to

$$\mathbf{E} \approx \frac{q\gamma\mathbf{R}_t}{4\pi\varepsilon_0\left[(\gamma\boldsymbol{\beta}\cdot\mathbf{R}_t)^2 + R_t^2\right]^{3/2}},$$
$$\mathbf{B} \approx \frac{\boldsymbol{\beta}\times\mathbf{E}}{c},$$
(S14)

where $R_t = |\mathbf{R}_t|$, with $\mathbf{R}_t = \mathbf{r}_i(t) - \mathbf{r}_j(t)$, and all right-hand-side variables are evaluated at time $t$. Calculating (S14) does not require explicit computation of the retarded time $t'$, and no history has to be maintained in the numerical solver. This algorithm has a time complexity of O($N^2$). As we note in Section S4, using (S14) yields results that closely approximate those obtained with (S13). This makes the formulation of space charge in (S14) suitable for investigations in our regime of interest, qualified by concerns related to the point of electron extraction noted in Section S3.

We note in passing that O($N\log N$) algorithms for space charge computation are available at the price of further approximations: ignoring relatively distant particles in the hierarchical method [S13], or assuming that all particles move at the same velocity in the method of multiple moments [S14].

S2.4    Radiation reaction and the Landau-Lifshitz equation

The radiation reaction term $R^\mu$ in (S4) accounts for the recoil the particle experiences when it radiates, in accordance with energy and momentum conservation laws. The effect of the recoil can be significant when energetic particles and high laser intensities are involved [S15]. The first generalization of radiation reaction to the relativistic case is the Lorentz-Abraham-Dirac (LAD) equation [S16, S17], which uses the term

$$R^\mu = \frac{\mu_0 q^2}{6\pi c}\left(\frac{d\alpha^\mu}{d\tau} - \frac{1}{c^2}\alpha^\nu \alpha_\nu \eta^\mu\right), \tag{S15}$$

where $\alpha^\mu = d\eta^\mu/d\tau$ is the 4-acceleration. However, the LAD equation suffers from physical inconsistencies such as the existence of ''runaway'' solutions, in which the particle acceleration diverges exponentially even in the absence of an external field [S4]. As a result, a more popular choice [S4] to model radiation reaction is the Landau-Lifshitz (LL) equation [S18], which uses the term

$$R^\mu = \frac{\mu_0 q^2}{6\pi c}\left[\frac{dF^{\mu\nu}}{d\tau}\eta_\nu + \frac{q^2}{m^2}F^{\mu\nu}F_\nu^{\ \alpha}\eta_\alpha + \frac{q^2}{m^2 c^2}\eta_\beta F^{\beta\nu}F_\nu^{\ \alpha}\eta_\alpha \eta^\mu\right], \tag{S16}$$

and avoids the nonphysical behavior of (S15). Although originally derived as an expansion of (S4) and (S15) to the first order in the fine structure constant, (S16) has since been separately obtained in a rigorous derivation using perturbation theory, and the LL equation has been argued to be a self-consistent equation of motion that is accurate for sufficiently small charged bodies with negligible dipole moments and spin [S19]. The LL equation has also been shown to be consistent with quantum electrodynamics to the order of the fine structure constant [S20, S21].

Our radiation reaction implementation uses (S16). Note that computing $dF^{\mu\nu}/d\tau$ requires knowledge of the partial derivatives in space and time of the electromagnetic fields, which we possess in analytical form since the fields are made up of analytical expressions (S6) and (S13) (or (S14)). Computing the partial derivative in time of (S13) would require knowledge of $d^2(\gamma\boldsymbol{\beta})/dt^2$ at the retarded time. Although (as mentioned in S1.3) we only store **r**, $\gamma\boldsymbol{\beta}$ and $d(\gamma\boldsymbol{\beta})/dt$, the additional term $d^2(\gamma\boldsymbol{\beta})/dt^2$ is readily approximated using the higher-order derivatives that have been pre-computed for the cubic spline interpolation method.

We have implemented radiation reaction in our routine to ensure as accurate a representation of the laser-driven multi-particle electrodynamics as possible. However, as we discuss in Section S4, the contribution of radiation reaction is negligible in all cases we consider in this work. Physically, this implies that the recoil the particle experiences when it emits radiation is not significant enough to affect any of its characteristics noticeably. Because of this, all results in our regime of interest may be accurately obtained just by solving Eq. (S4) with $R^\mu = 0$. This confirms that radiation reaction neither significantly affects nor is responsible for the physical effects studied in this work.

## S3  Linear acceleration with more energetic laser pulses

In this section, we present results corresponding to Figs. 2 and 4 of the main text for larger laser pulse energies and electron pulse charges. Specifically, we demonstrate the acceleration of 30 keV electrons (2.5% spread) to 61 MeV (0.5% energy spread; see Fig. S3) and 205 MeV (0.25% energy spread; see Fig. S4) using 250 mJ and 2.5 J lasers respectively. Increasing the amount of charge in the electron pulse from -0.2 fC to -2 fC in the 2.5 J laser case results in a different optimization point – where the optimization procedure and figure-of-merit is as delineated in the main text – such that the final energy is now 217 MeV (0.68% energy spread; see Fig. S5). Unless otherwise specified, all other laser and electron pulse parameters are the same as in Fig. 1 of the main text.

Like Fig. 4 of the main text, Figs. S6, S7 and S8 emphasize that for a reasonably wide range of parameters, one obtains a final electron pulse with significant energy gain, low emittance and low energy spread. Consequently, our results presented in Figs. S3, S4 and S5 are not sensitive to small fluctuations from specific choices of parameters. This is further emphasized by additional comprehensive simulations summarized in Fig. 3 of the main text. That figure shows that for a wide range of laser and electron pulse parameters, the normalized trace-space emittance falls in the few nm-rad range, the energy spread falls in the few-percent range, and the mean electron kinetic energy attains substantially relativistic values.

Unlike the parameters used in Fig. 1, which are already achievable in leading laser facilities, these examples serve mainly to motivate technological improvements towards Joule-level, few-cycle laser pulses, some of which is already underway in the form of the Petawatt Field Synthesizer [S22] and the Extreme Light Infrastructure [S23]. An exploration of different types of electromagnetic field configurations is likely to yield more impressive acceleration performance with more modest laser properties [S24].

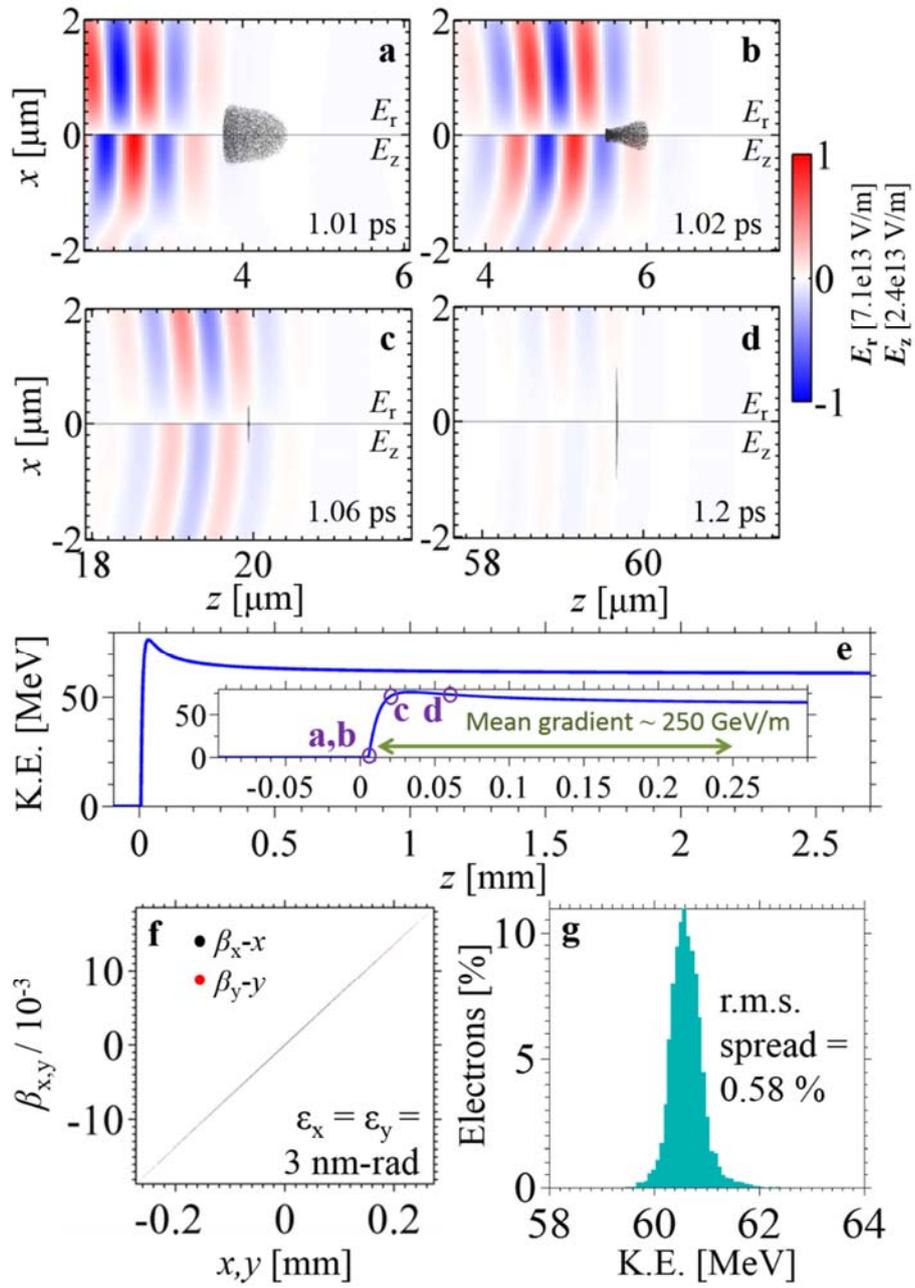

**Figure S3.** Same as Fig. 2 of the main text, except the laser pulse energy is now 250 mJ.

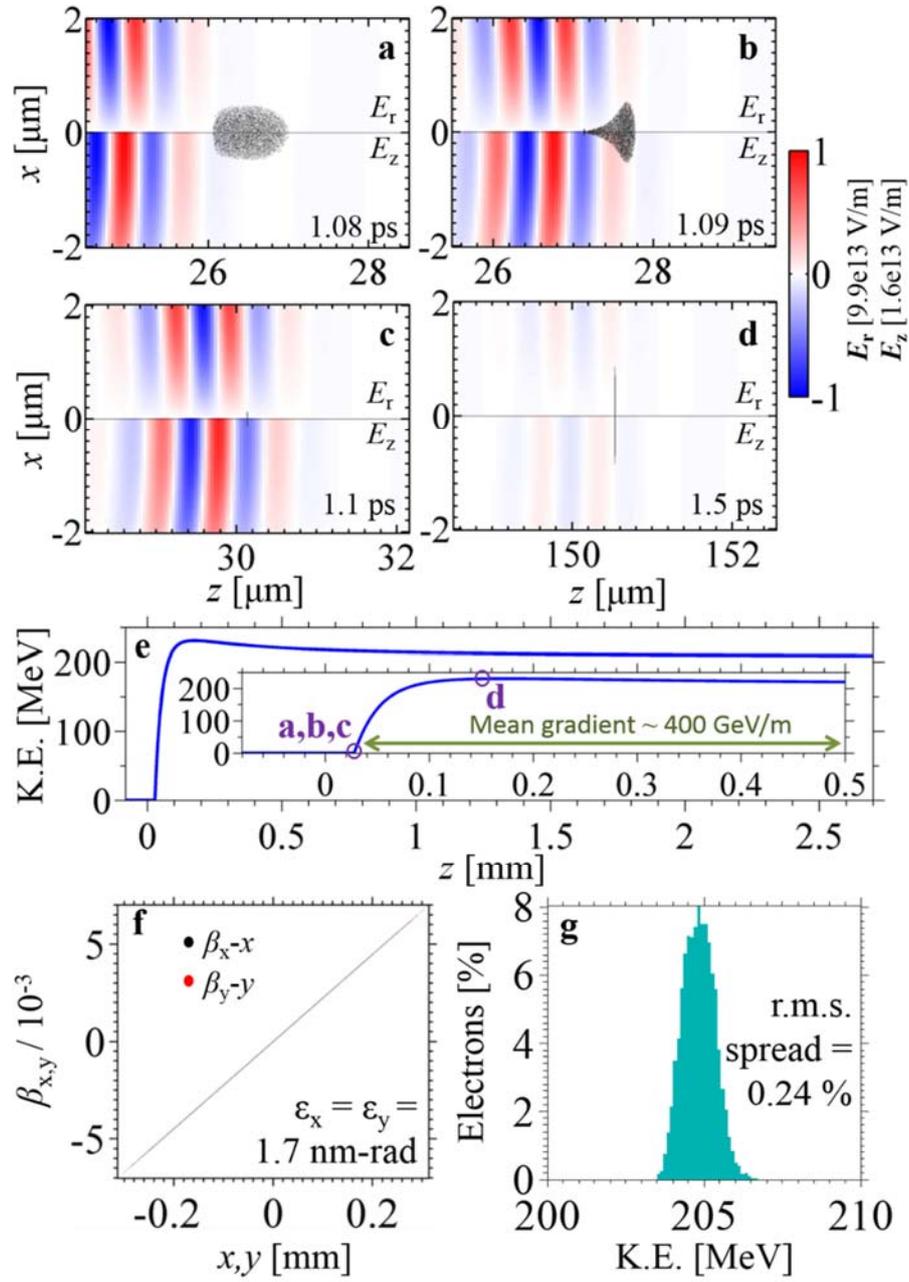

**Figure S4.** Same as Fig. 2 of the main text, except the laser pulse energy is now 2.5 J.

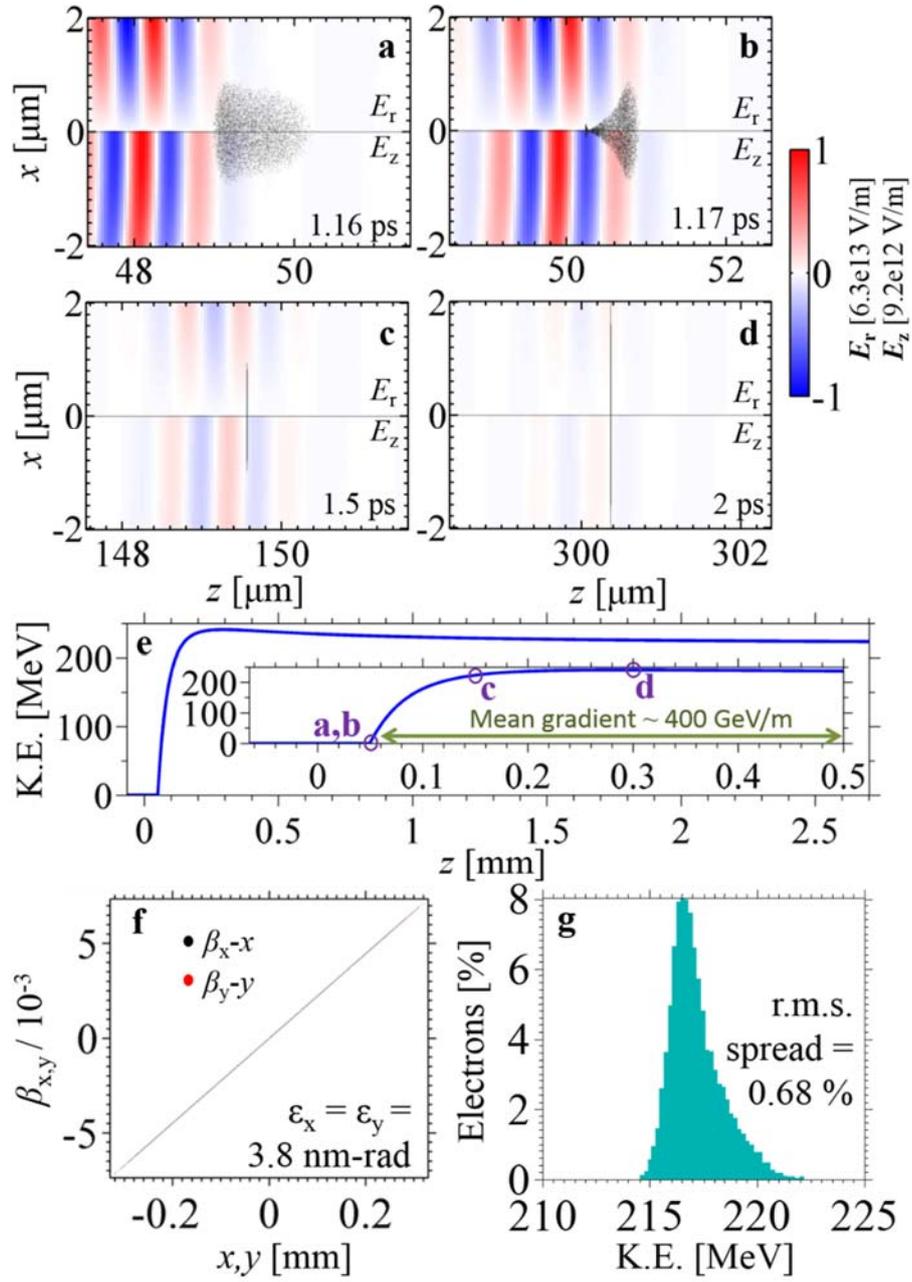

**Figure S5.** Same as Fig. 2 of the main text, except the laser pulse energy is now 2.5 J and the electron pulse contains -2 fC of charge.

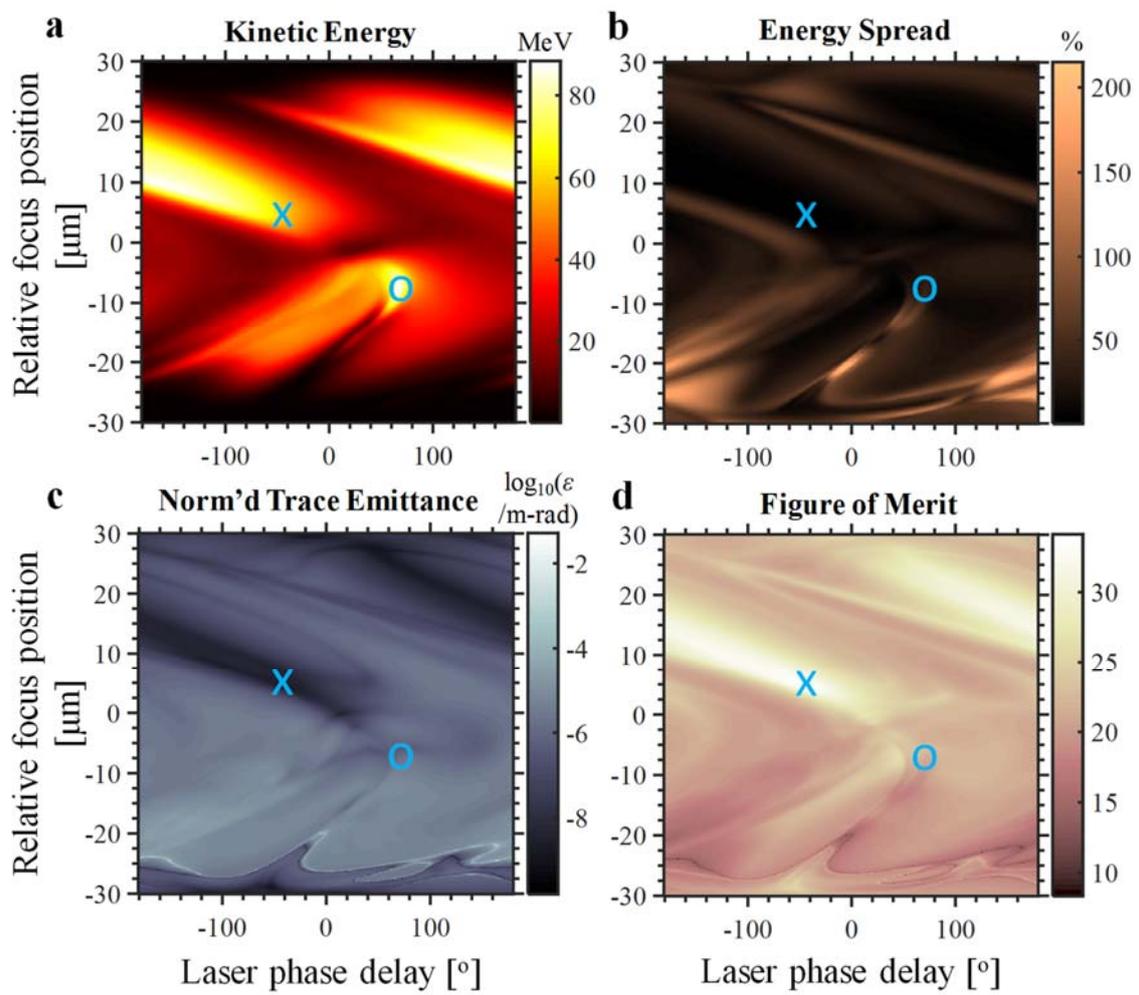

Figure S6. Same as Fig. 4 of the main text, except the laser pulse energy is now 250 mJ. The optimal solution corresponds to the results in Fig. S3.

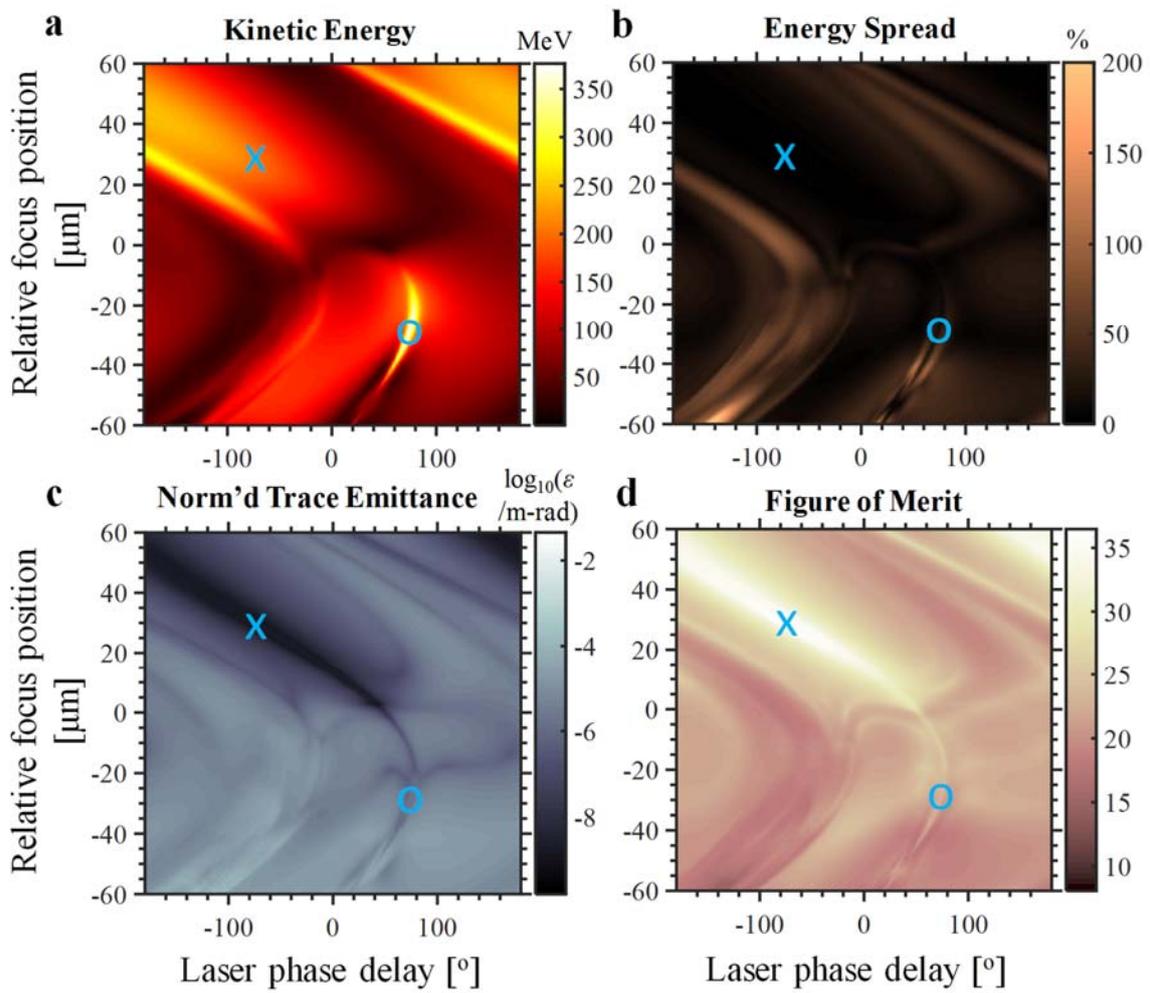

**Figure S7. Same as Fig. 4 of the main text, except the laser pulse energy is now 2.5 J. The optimal solution corresponds to the results in Fig. S4.**

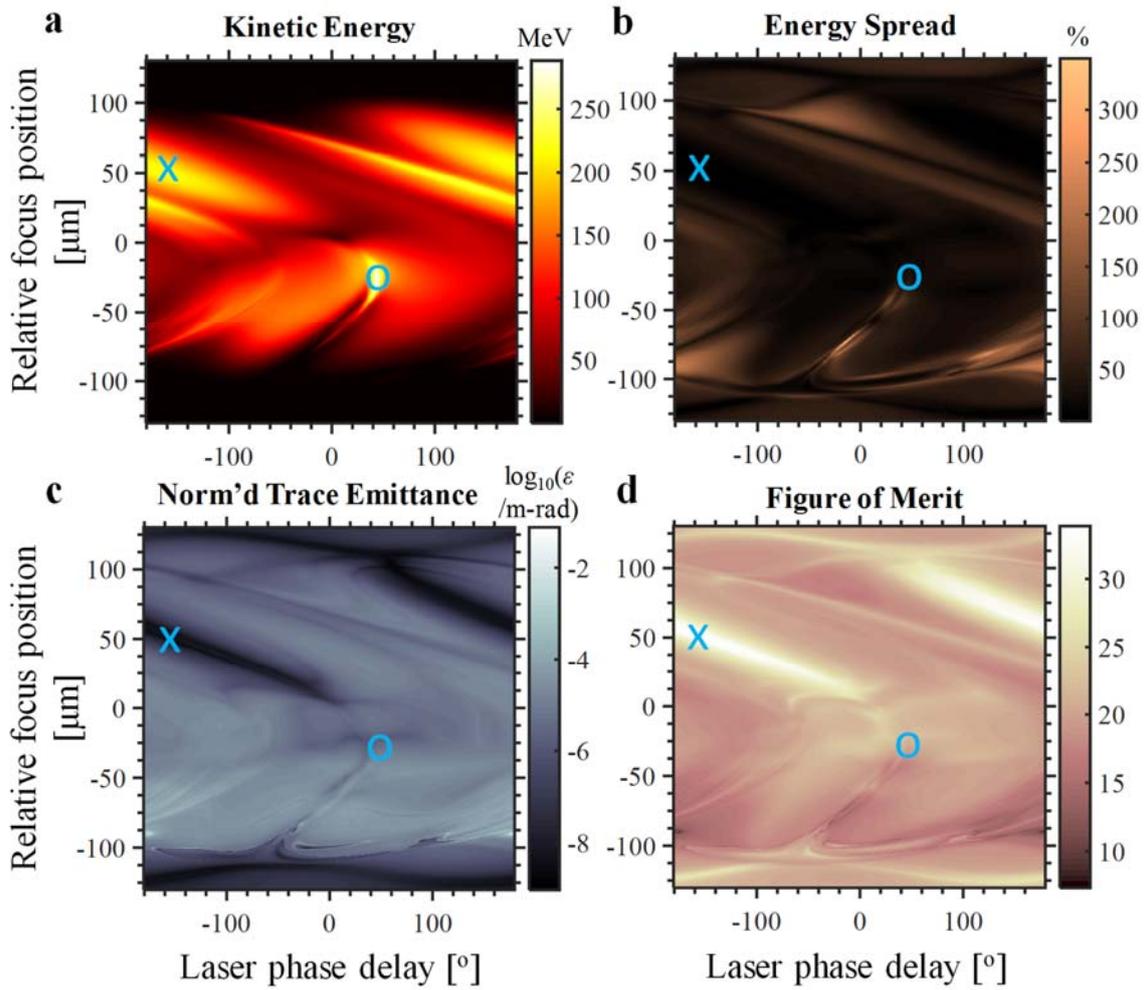

**Figure S8.** Same as Fig. 4 of the main text, except the laser pulse energy is now 2.5 J and the electron pulse contains -2 fC of charge. The optimal solution corresponds to the results in Fig. S5.

## S4    Evolution of electron statistics during linear acceleration in free space

In this section, we take a closer look at the statistical evolution of the electron pulse in the course of the laser-driven electron acceleration process. We illustrate the effects of the far field and radiation reaction – which were both included in our simulations – with four cases corresponding to laser pulse energies and electron pulse charges of 25 mJ and -0.2 fC (Fig. S9); 250 mJ and -0.2 fC (Fig. S10); 2.5 J and -0.2 fC (Fig. S11); and 2.5 J and 2 fC (Fig. S12). Unless otherwise stated, all laser and electron pulse parameters are the same as – or determined via the same optimization procedure as – those corresponding to Figs. 1 and 2 of the main text.

The "oscillations" seen in the emittance (Figs. S9b, S10b, S11b and S12b) for a period of time after the main interaction at 1 ps is a result of the now-relativistic electrons interacting with the tail of the laser pulse. These "oscillations" continue long after the kinetic energy and energy spread have mostly reached their steady state (~2 ps), and last for durations between 4 ps (Fig. S9b) and 2ns (Fig. S12b). Correspondingly, the period of such "oscillations" is also relatively large, as expected from the fact that the final electrons are relativistic and travel at a velocity very close to the velocity of light in free space, making it unsurprising that some weak interaction persists at the tail of the laser pulse for an extended period of time. Note, however, that this weak interaction has a negligible effect on the final kinetic energy and energy spread of the electron pulse.

Radiation reaction and the far field affect the electron pulse behavior negligibly at sub-Joule laser energies (Figs. S9 and S10). At Joule-level laser energies (Figs. S11 and S12), radiation reaction continues to play a negligible role. The far-field affects the energy spread, emittance and standard deviation by a noticeable amount over very long interaction times (many ns). However, practical considerations will typically limit this interaction by extracting the electrons or blocking the laser, making this long time interaction less significant: e.g., by a bending magnet, or by injection into a second stage whose fields overwhelm the effects of the far field propagating with the electron pulse. Extracting the electron bunch would yield better emittance if performed close to one of the emittance minima (about 30 cm in length) or before the oscillations (about 5cm in length) when a shorter distance is desired (see Figs. S11b, S12b). We have done this in obtaining the results presented in the main text. Otherwise, as Figs. S11b and S12b show, the far field can cause a steady degradation of the emittance quality. Note, however, that the far field has negligible effect on the mean kinetic energy and energy spread, at least up to a meter of propagation. Fig. S12b shows that this effect of the far field is exacerbated when the charge density is increased. As described in Section S2.3, we have modeled the far field exactly using the Liénard-Wiechert potentials. For all cases studied, we have also verified that the near fields of the Liénard-Wiechert potentials are extremely well-approximated by the Lorentz-boosting procedure widely adopted in particle-tracking software (e.g., GPT). This implies that standard particle-tracking software would yield reasonably good estimates for a short time after the interaction, but would not predict the gradual long time degradation due to the far field interaction because the far field and radiation reaction are not typically taken into account in standard particle-tracking software for accelerator systems (e.g., PARMELA, GPT).

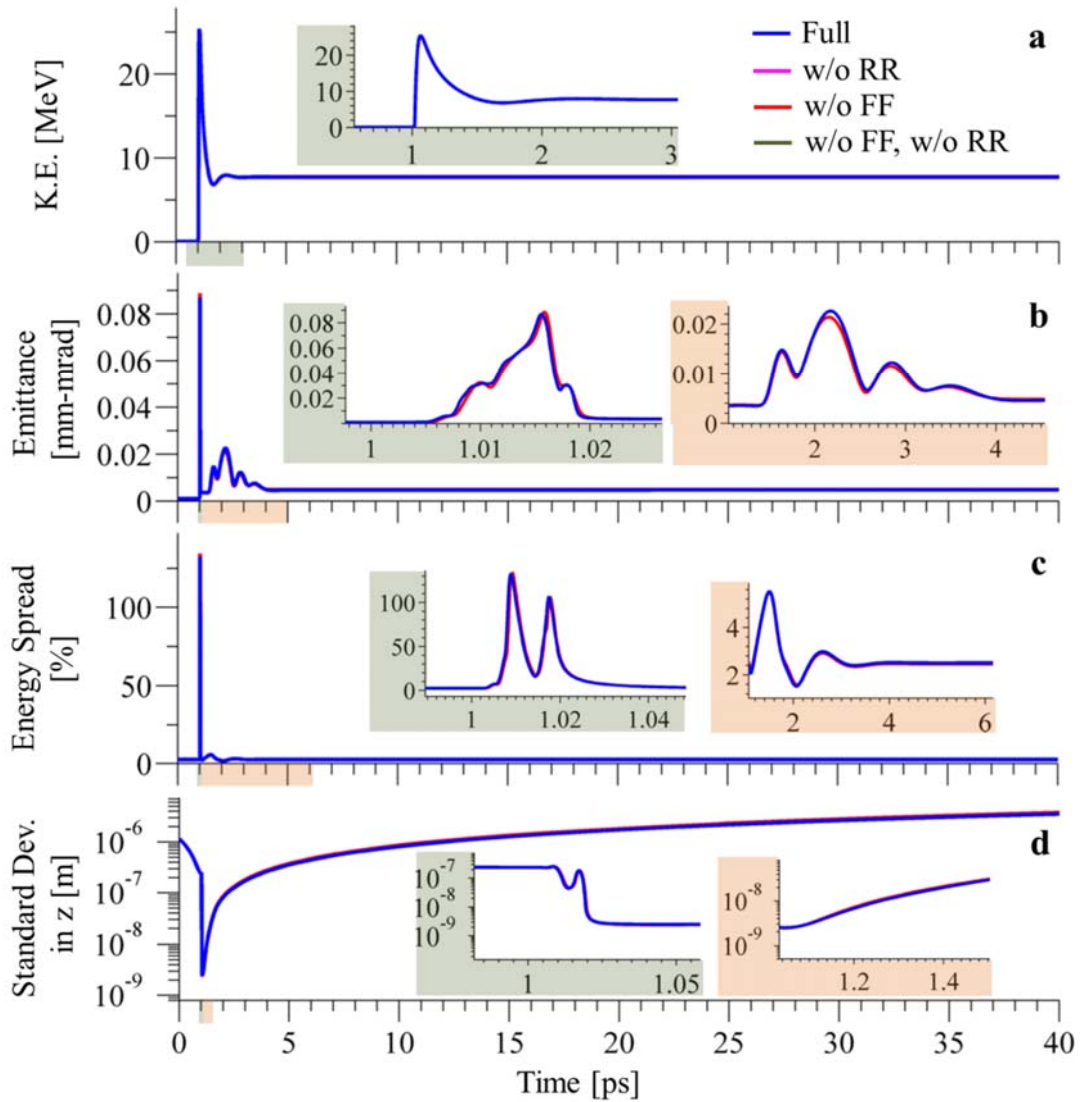

**Figure S9.** Evolution as a function of time of the electron pulses (a) mean kinetic energy, (b) normalized trace-space emittance, (c) relative energy spread, and (d) standard deviation, for the case in Figs. 1 and 2 of the main text (25 mJ laser energy, -0.2 fC charge). "Full": Simulation of the exact physical problem with radiation reaction (captures both near- and far-field effects in inter-particle interactions, as well as the effect of radiation reaction). "w/o RR": Same as "Full" but without including radiation reaction. "w/o FF": Same as "Full" but without considering the far-field in inter-particle interactions. "w/o RR, w/o FF": Simulation

captures only the near-field in inter-particle interactions, and does not consider radiation reaction. In all plots, magenta lines ("w/o RR") practically lie under blue lines ("Full"), and dark green lines ("w/o FF, w/o RR") practically lie under red lines ("w/o FF").

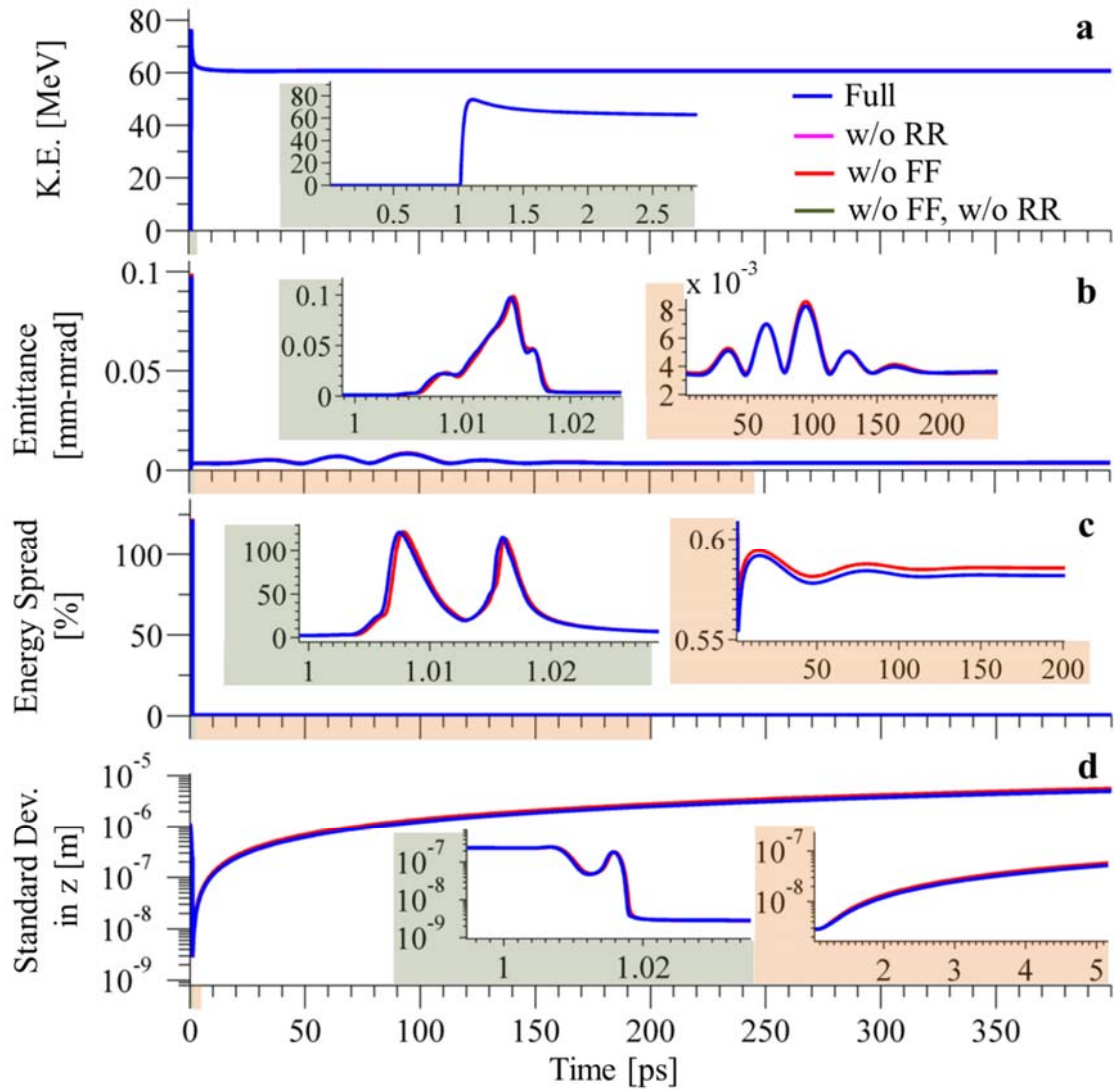

**Figure S10. Same as Fig. S9 but for the case in Fig. S3 (250 mJ laser energy, -0.2 fC charge). In all plots, magenta lines ("w/o RR") practically lie under blue lines ("Full"), and dark green lines ("w/o FF, w/o RR") practically lie under red lines ("w/o FF").**

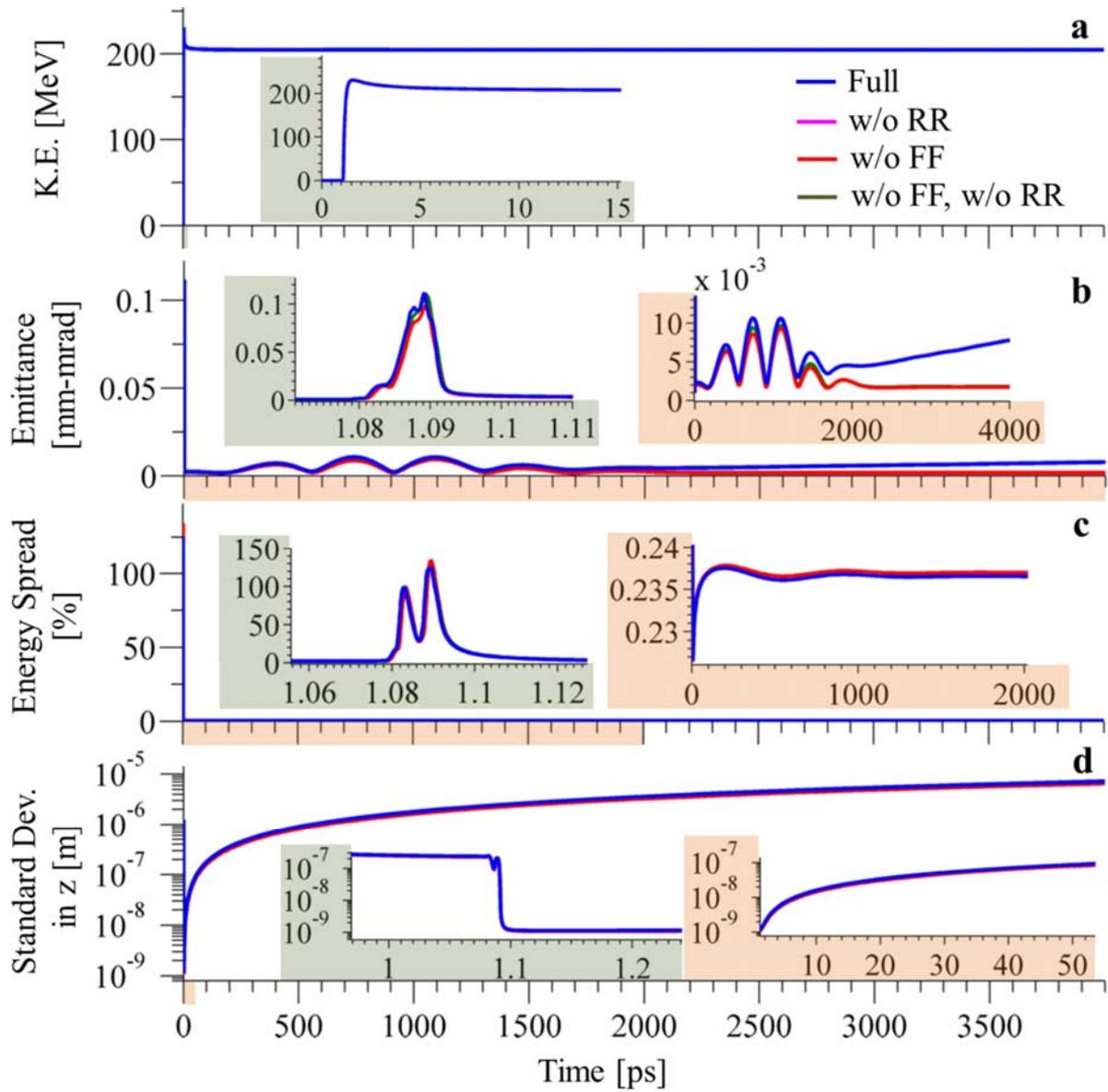

**Figure S11.** Same as Fig. S9 but for the case in Fig. S4 (2.5 J laser energy, -0.2 fC charge). In all plots, magenta lines ("w/o RR") practically lie under blue lines ("Full"), and dark green lines ("w/o FF, w/o RR") practically lie under red lines ("w/o FF").

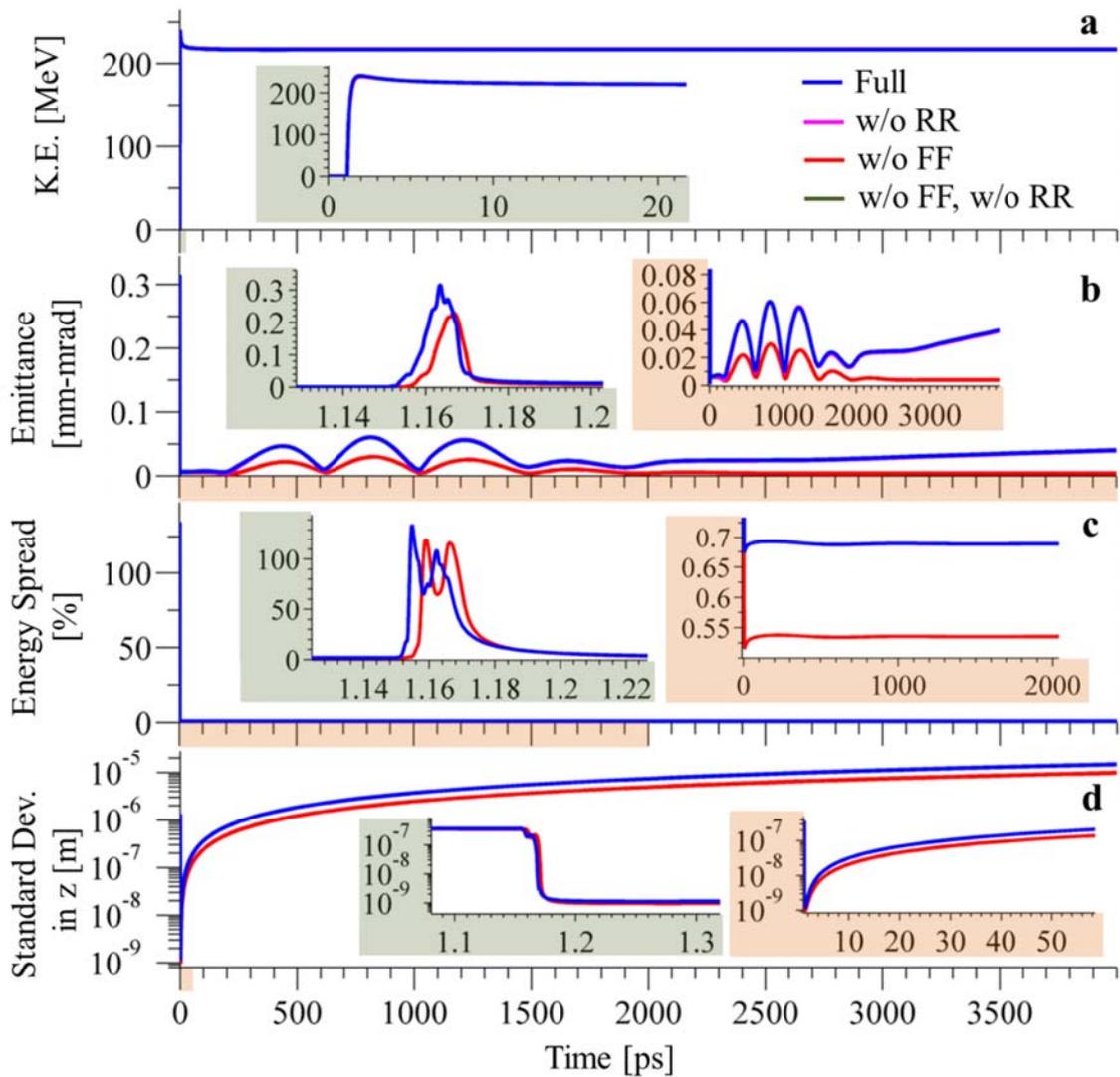

**Figure S12.** Same as Fig. S9 but for the case in Fig. S5 (2.5 J laser energy, -2 fC charge). In all plots, magenta lines ("w/o RR") practically lie under blue lines ("Full"), and dark green lines ("w/o FF, w/o RR") practically lie under red lines ("w/o FF").

## S5    High-quality acceleration for substantial initial electron energy spreads

In this section, we explore the impact of larger initial electron energy spreads for the cases studied in Fig. 3 of the main text. From Figs. S13, S14, S15 and S16, we see that doubling the initial energy spread from 2.5% to 5% has practically negligible effect on the scheme's performance, over a wide range of parameters. In fact, an initial energy spread as large as 10% can still produce final trace-space emittances in the nm-rad range for all cases (distinguished by color) in Fig. 3 of the main text. Figs. S15 and S16 show that the scheme is capable of producing quasi-monoenergetic, highly-relativistic output (e.g., a 200 MeV bunch with < 2% energy spread) from initial bunches of energy spreads as large as 40%. These results imply that relatively large initial energy spreads can be accommodated by our scheme with relatively small deterioration in output quality.

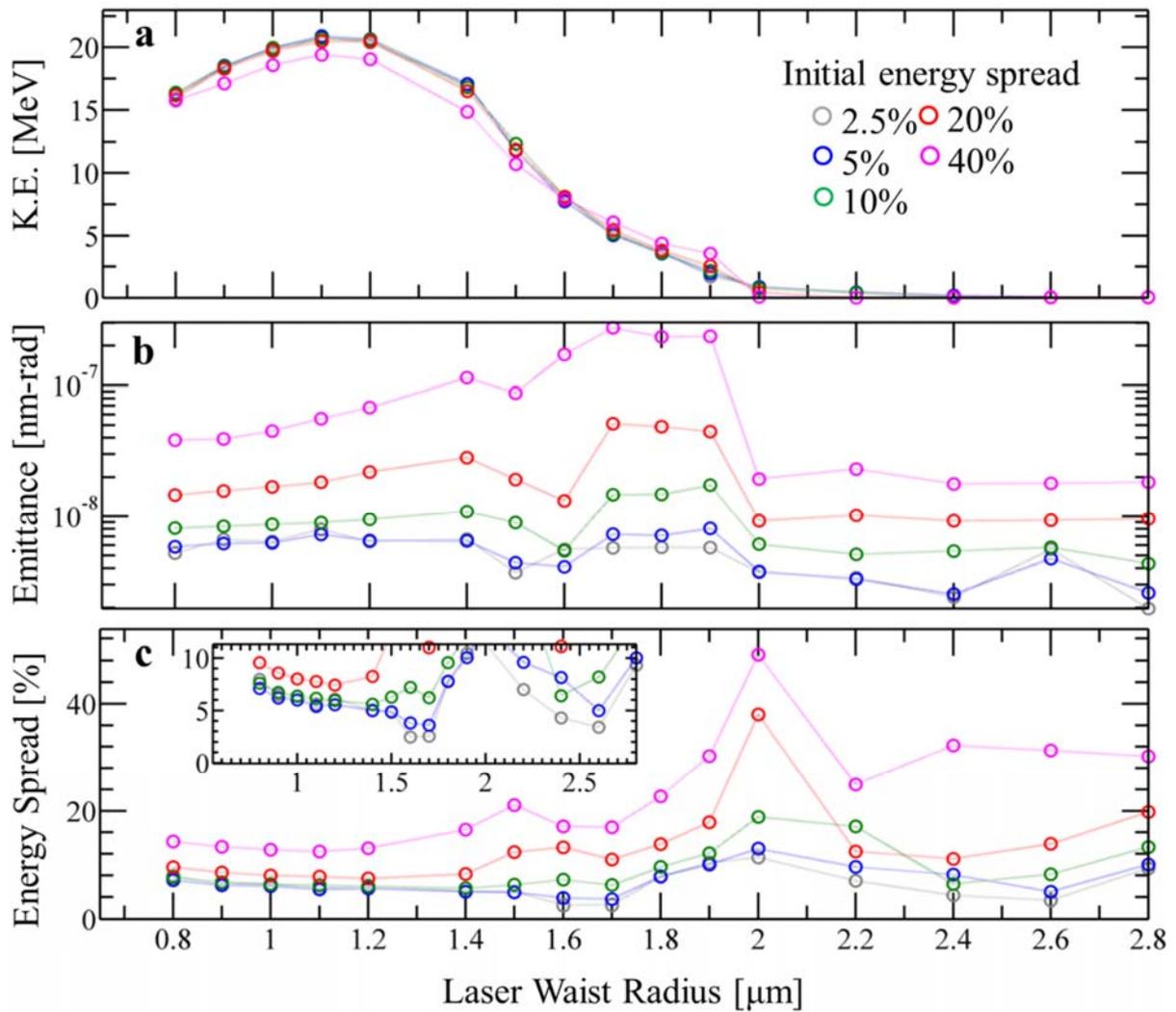

**Figure S13.** Same as Fig. 3 in the main text for several values of electron energy spread between 2.5% and 40%. The laser pulse energy and total charge are fixed at 25 mJ and -0.2 fC respectively here. These results imply that relatively large initial energy spreads can be accommodated by our scheme with relatively small deterioration in output quality.

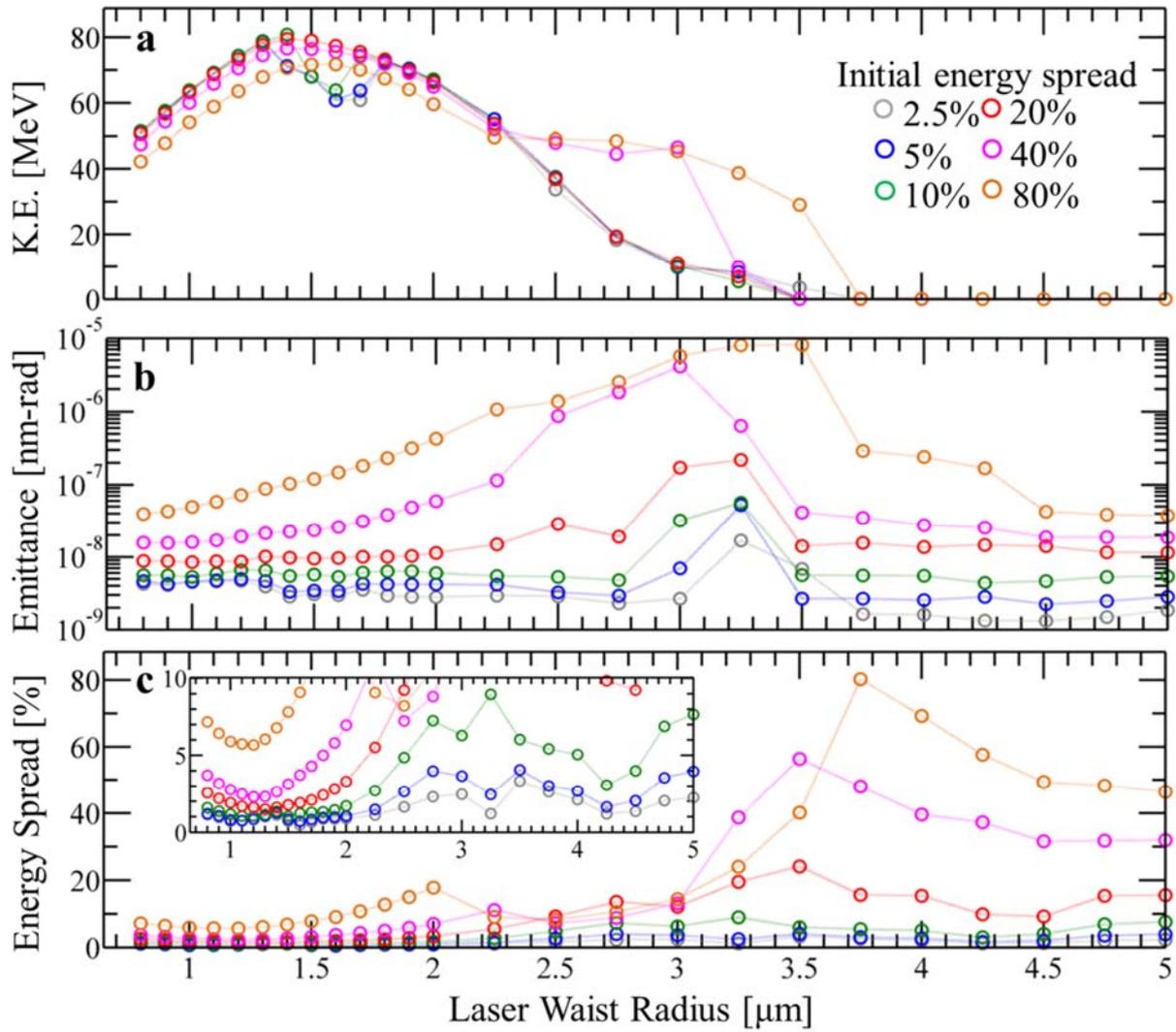

**Figure S14.** Same as Fig. S13 but for the case of 250 mJ laser energy, -0.2 fC charge, and for several options of initial energy spreads up to 80%.

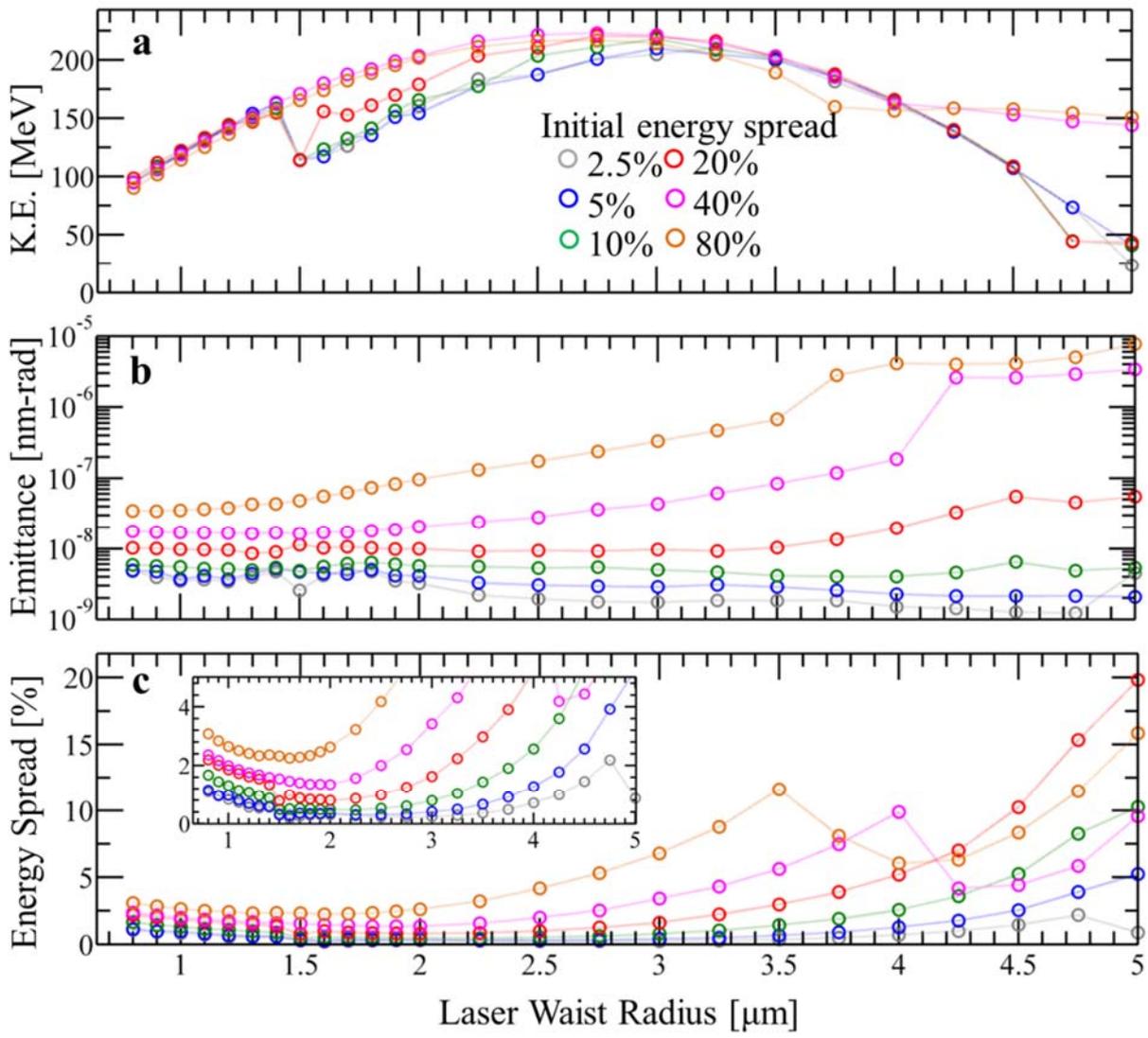

**Figure S15.** Same as Fig. S14 but for the case of 2.5 J laser energy, -0.2 fC charge.

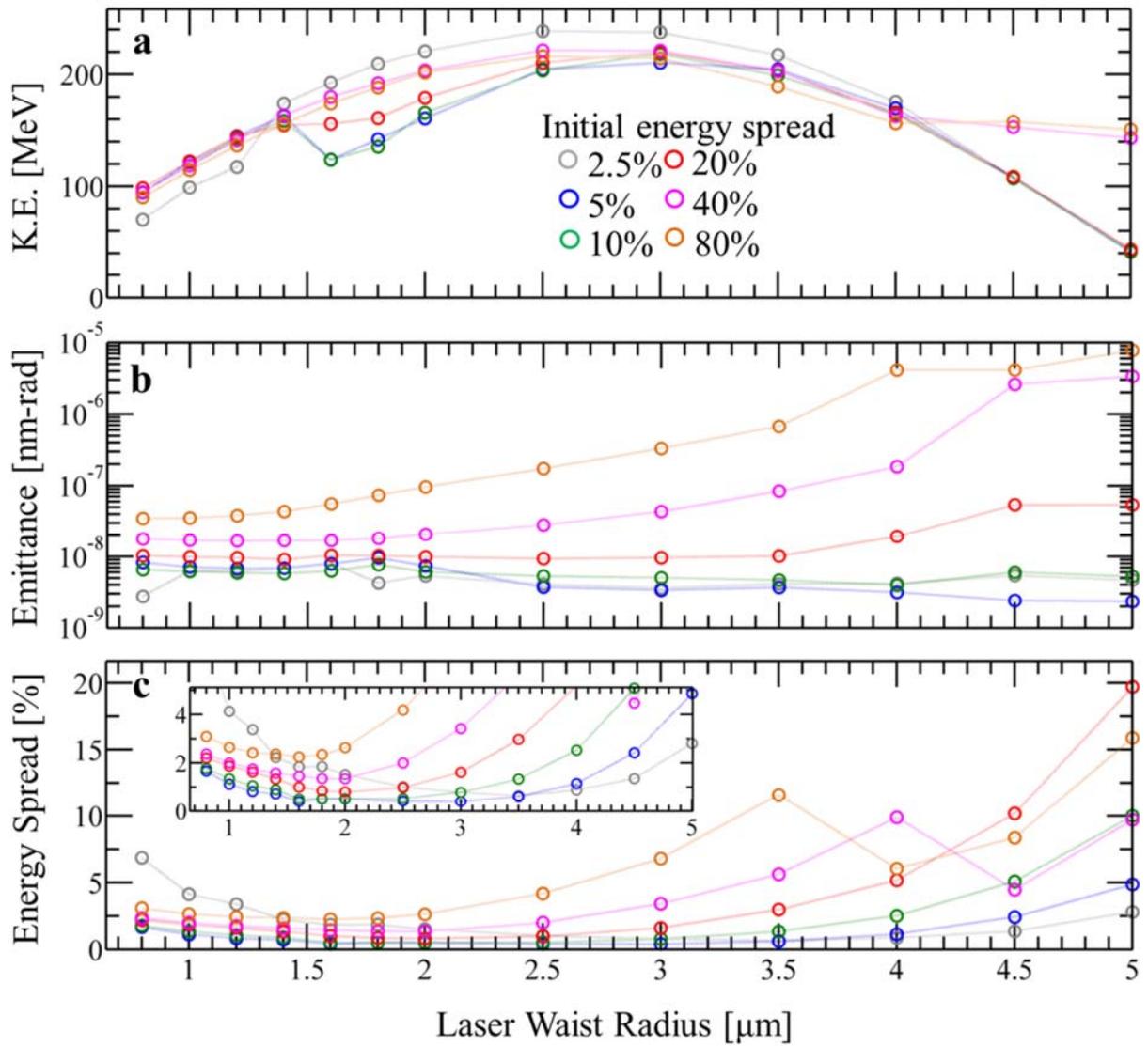

Figure S16. Same as Fig. S14 but for the case of 2.5 J laser energy, -2 fC charge.

## S6   Linear acceleration with 6 fs laser pulses

In this section, we study the performance of the scheme using 6 fs (intensity FWHM) laser pulses, which are more readily available compared to the 3 fs pulses explored in the main text.

The results, obtained under the same conditions and in the same fashion as Fig. 3 in the main text, are shown in Fig. S17. We note that an initial electron pulse of kinetic energy 30 keV (2.5% energy spread) and -0.2 fC of charge can be accelerated to a final kinetic energy of 2.7 MeV (3.2% spread) by with a laser pulse energy of 25 mJ; and to a final kinetic energy of 103 MeV (0.5% spread) by a laser pulse energy of 2.5 J. If the initial charge is increased to -2 fC, a final energy of 95 MeV (1.8% spread) can be obtained. Thus, we see that even when the laser pulse duration is increased from 3 fs to 6 fs, the scheme is still capable of substantial and high-quality acceleration of multi-electron pulses.

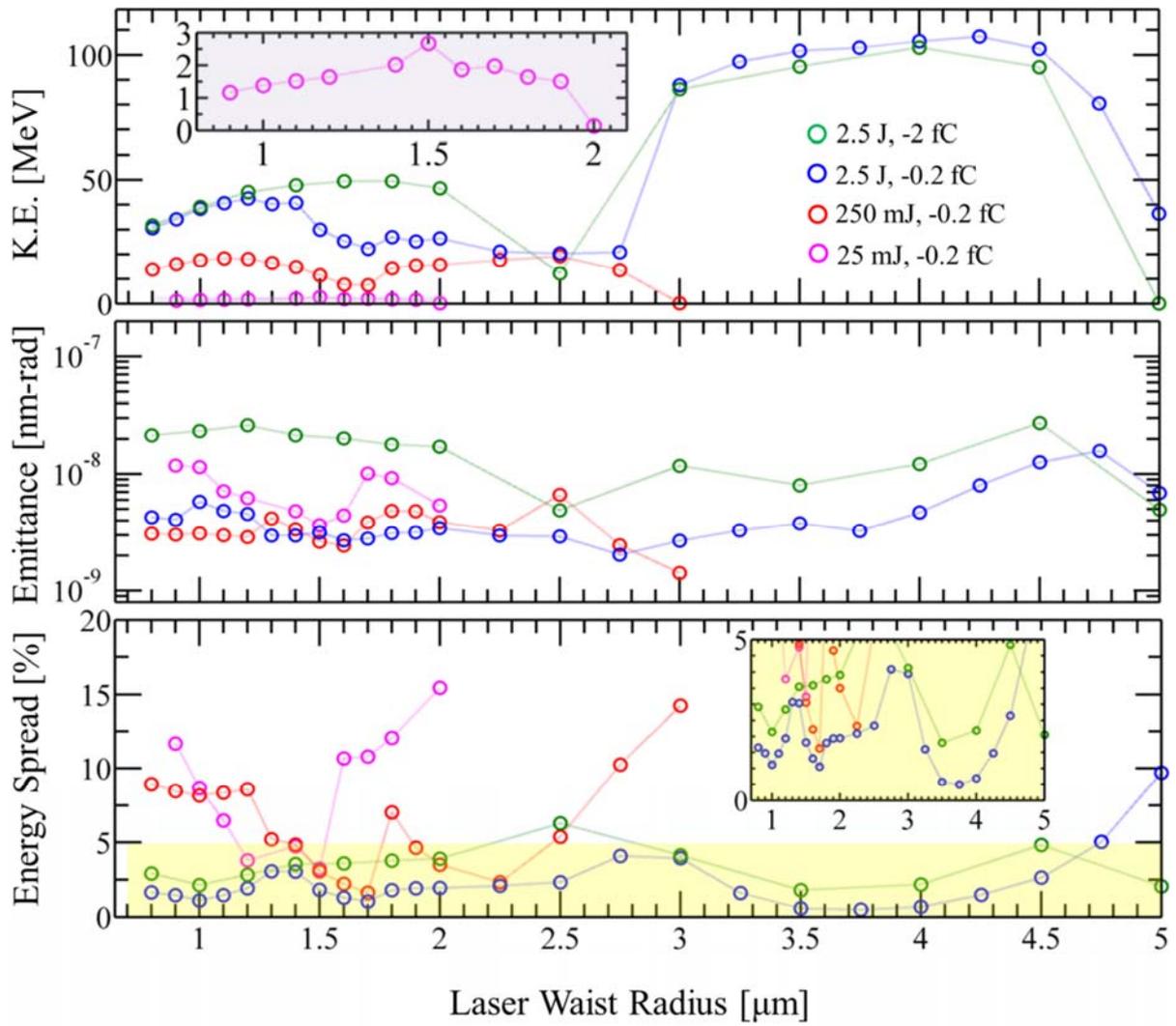

Figure S17. Same as Fig. 3 but for a 6 fs laser pulse.